\documentclass[a4paper,11pt]{article}
\pdfoutput=1

\usepackage{jheppub}

\usepackage[T1]{fontenc}

\usepackage{graphicx,subcaption}
\usepackage[export]{adjustbox}

\usepackage{psfrag}

\usepackage[percent]{overpic}
\usepackage[dvipsnames]{xcolor}

\usepackage{slashed}

\usepackage{comment}

\usepackage{placeins}

\usepackage{float}

\usepackage{fontawesome5}

\newcommand{\be}{\begin{equation}}
\newcommand{\ee}{\end{equation}}
\newcommand{\bea}{\begin{eqnarray}}
\newcommand{\eea}{\end{eqnarray}}

\newcommand{\uv}{\rm UV}
\newcommand{\ir}{\rm IR}
\newcommand{\T}{T^{\mu \nu}}
\newcommand{\ZIR}{\mathbf{Z}_{\rm IR}}

\newcommand{\GIR}{\mathbf{\Gamma}_{\rm IR}}

\newcommand{\U}{\mathrm{U}}
\newcommand{\SU}{\mathrm{SU}}

\newcommand{\suc}{SU(N_1)}
\newcommand{\suw}{SU(N_2)}

\newcommand{\ie}{\emph{i.e.}~}
\newcommand{\eg}{\emph{e.g.}~}

\DeclareRobustCommand{\Sec}[1]{Sec.~\ref{#1}}

\DeclareRobustCommand{\App}[1]{App.~\ref{#1}}
\DeclareRobustCommand{\Tab}[1]{Table~\ref{#1}}

\DeclareRobustCommand{\Fig}[1]{Fig.~\ref{#1}}

\DeclareRobustCommand{\Eq}[1]{Eq.~(\ref{#1})}

\newcommand{\Lag}{\mathcal{L}}

\newcommand{\github}[1]{%
   \href{#1}{\faGithubSquare}%
}

\title{\boldmath Infrared Anomalous Dimensions at Three-Loop in the SM from Conserved Currents}

\author[a,b]{Michael Stadlbauer,}
\author[a]{Tobias Theil}

\affiliation[a]{Technical University of Munich, TUM School of Natural Sciences, Physics Department, James-Franck-Str. 1, 85748 Garching, Germany}
\affiliation[b]{Max Planck Institute for Physics, Boltzmannstr. 8, 85748 Garching, Germany}

\emailAdd{michael.stadlbauer@tum.de}
\emailAdd{tobias.theil@tum.de}

\abstract{We study the infrared structure of the Standard Model (SM) restricted to the first generation of fermions, including the full SM gauge group, up to three-loop order, and determine the resulting cusp and collinear anomalous dimensions for all gauge groups and particles of the theory. We observe that starting from three loops, the resulting cusp anomalous dimensions include terms involving two distinct couplings, contrary to previous claims in the literature. We give a detailed explanation of this observation and explore the origin of these terms.
We provide a supplementary file where all infrared anomalous dimensions calculated in this work are collected~\href{https://github.com/michael-stadlbauer/AnomalousDimensions.git}{\faGithub}. Our results are consistent with the existing literature wherever comparisons are possible.}

\preprint{TUM-HEP-1527/24}

\begin{document} 
\maketitle
\flushbottom

\section{Introduction}

It is well known that loop amplitudes exhibit both ultraviolet (UV) and infrared (IR) divergences. While the former can be treated with renormalization techniques, the latter cancel only in physical observables, such as the total cross section, when combined with contributions from real emission diagrams. Consequently, after renormalization, the surviving poles of loop amplitudes can be related to the divergences arising upon phase space integration over real radiation processes. Furthermore, they are tightly connected to large logarithms that, in the case of large scale separations, can spoil the perturbative expansion and need to be resummed. A systematic resummation procedure has been developed in the context of the soft-collinear effective theory \cite{Bauer:2000yr,Bauer:2001ct, Bauer:2001yt, Bauer:2000ew, Beneke:2002ph, Beneke:2002ni}. Within this framework, the operator coefficients obey a renormalization group equation. In fact, the corresponding anomalous dimensions, which are constructed from two constants, the cusp and collinear anomalous dimension, can be directly computed from the infrared poles of usual scattering amplitudes

In this paper, we aim to compute the cusp and collinear anomalous dimensions within the Standard Model restricted to only the first generation of fermions up to the three-loop order, using form factors of conserved currents with two external states. These have two beneficial properties: First, they are the simplest objects that exhibit a non-trivial structure of IR divergences. Second, after coupling renormalization, the currents themselves are UV finite, such that the extraction of the IR poles becomes trivial. Within QCD such form factors are known up to three loops for some years now \cite{Matsuura:1988sm,Matsuura:1987wt,Kramer:1986sg,Harlander:2000mg,Gehrmann:2005pd,Moch:2005tm,Gehrmann:2010ue}, where a vector current with no association to the gauge group has been used for the fermions. For gluons, on the other hand, no easy form factor exists within the theory. Therefore, most authors resort to the addition of an auxiliary coupling to an external scalar via a dimension five operator. This, however, is not UV finite, and to extract the IR poles, this new operator first has to be UV renormalized at the same loop order. Rather recently, these form factors have been computed at four loops in QCD \cite{vonManteuffel:2020vjv}.

While we will follow a very similar route, the main difference is that we use the energy-momentum tensor $T_{\mu\nu}$ as the conserved current of our choice.
A similar approach has been taken in \cite{Baratella:2022nog} for the Yukawa sector using on-shell amplitude methods as well as in \cite{deFlorian:2013sza, Ahmed:2015qia}, where the authors however restrict their theory to the QCD sector of the Standard Model.
Since $T_{\mu\nu}$ couples to all particles in the theory universally, it can be used to compute IR poles for all particles in exactly the same way. And because it is UV finite, we need to only include UV renormalization of the couplings up to two loops.\footnote{In addition, if there would be any particle not charged under any of the gauge groups in the theory, as \eg a right-handed neutrino, $T_{\mu\nu}$ could, in principle, still be used to extract the IR structure.}

We will use the remainder of this section to illustrate the structure of IR divergences for general amplitudes and form factors. In addition, we present the properties of conserved currents that make them exceptional objects for studying IR divergences in more detail. In section \Sec{sec:Model}, we define our model and outline our strategy to compute the UV counterterms within the model. The corresponding beta functions can be found in the supplementary file \texttt{BetaFunc\_SM\_2Loop.m}. 
The method used to compute the form factors up to the three-loop order is explained in \Sec{sec:computation}.
The results of our computation are discussed in \Sec{sec:results}, focusing on mixed contributions that were previously thought to arise only at the four-loop order for the first time. Selected parts of our results can be found in \App{app:gammas}, and all results are summarized in the supplementary file \texttt{IRAnomDim\_SM\_3Loop.m}. 
We draw our conclusions in \Sec{sec:conclusion}.

\subsection{Structure of IR Divergences} \label{sec:IR_struc}

We define the amputated and renormalized $n$-point Green's function of the fields $\Phi_i$ with an operator $\mathcal{O}$ inserted by
\begin{equation}
    \langle \mathcal{O} \rangle \equiv \langle \Omega | T \{ \mathcal{O}(q) \Phi_{i_1}(p_1) \ldots \Phi_{i_n}(p_n) \} | \Omega \rangle.
\end{equation}
For $\mathcal{O} = \mathbf{1}$, we obtain the usual Green's function without sources, which, via the LSZ formula, is directly related to the S-matrix.
The bare Green's functions $\langle \mathcal{O} \rangle_0$ is related to the renormalized one via
\begin{equation}
    \langle \mathcal{O} \rangle = Z_{G}^{-1} \left. \langle \mathcal{O} \rangle_0 \right|_{g^{(0)}_i \rightarrow g_i},
\end{equation}
where the coupling renormalization is implicitly realized by expressing the bare couplings $g^{(0)}_i$ in terms of the renormalized ones $g_i$ on the RHS of the equation. We include in the constant $Z_{G}$ the renormalization of the external fields, the $\uv$ renormalization of the operator itself $Z^\mathcal{O}_{\rm UV}$ and a factor $\hat{\mathbf{Z}}_{\rm IR}$ to remove any $\ir$ divergences appearing at any given loop order \ie
\begin{equation}
    Z_G = \prod_{\Phi_i} Z_{\Phi_i}^{-n_i/2} Z^\mathcal{O}_{\rm UV} \hat{\mathbf{Z}}_{\rm IR} = Z^\mathcal{O}_{\rm UV} \mathbf{Z}_{\rm IR},
\end{equation}
where in the last expression, we collect the external field renormalization constants and $\hat{\mathbf{Z}}_{\rm IR}$ to define the $\ir$ renormalization factor $\mathbf{Z}_{\rm IR}$. \\

This $\ir$ renormalization factor $\mathbf{Z}_{\rm IR}$ is universal as it is the same for any Green's function with the same set of $n$ external particles and its structure has been extensively studied in literature \cite{Becher:2009cu,Becher:2009qa,Catani:1996jh,Catani:1996vz,Catani:1998bh}. \\

The $\ir$ anomalous dimension of a general $n$-point amplitude $\mathbf{\Gamma}_{\rm IR}$ is defined via $\mathbf{Z}_{\rm IR}$ by
\begin{equation} \label{eq:GIR_def}
    \mathbf{\Gamma}_{\rm IR} = - \mathbf{Z}_{\rm IR}^{-1} \mu \frac{d \mathbf{Z}_{\rm IR}}{d \mu},
\end{equation}
and is, up to three loops, always of the form
\begin{equation} \label{eq:IR_anom_dim_general}
    \boldsymbol{\Gamma}_{\rm IR}= \sum_{\{G\}} \sum_{(i, j)} \frac{\boldsymbol{T}^G_i \cdot \boldsymbol{T}^G_j}{2} \gamma^G_{\rm {cusp}} \log \left( \frac{\mu^2}{-s_{i j}} \right) + \sum_i \gamma^i_{\rm coll},
\end{equation}
where the sum over $\{G\}$ runs over all gauge groups involved, the sums over $i$ and $j$ are over unordered tuples of particle indices, which determines the representation of the generator $\boldsymbol{T}$ and $s_{ij} = (p_i+p_j)^2$. Note that the generator of a particle not charged under a certain gauge group is zero and, hence, does not contribute to the above sum. When acting on a gauge group singlet state, which will always be the case in this work, the sum over the generators simplifies to
\begin{equation} \label{eq:rel_gen_casimir}
    \sum_{(i, j)} \boldsymbol{T}_i^G \cdot \boldsymbol{T}_j^G = -\sum_i (\boldsymbol{T}_i^G)^2=-\sum_i C_i^G,
\end{equation}
with $C_i^G$ being the quadratic Casimir operator.
The operator formula in \Eq{eq:IR_anom_dim_general} captures the universallity of IR divergencies. Once $\gamma_{\rm cusp}$ and $\gamma_{\rm coll}$ are known up to a given loop order, \Eq{eq:IR_anom_dim_general} can be used to determine the renormalization factor $\mathbf{Z}_{\ir}$ for a given configuration of external states.
At this point, we want to stress that as long as \Eq{eq:IR_anom_dim_general} holds,
the dependence on the species of external particles enters the logarithmic term only through the quadratic Casimir operators. In other words, these terms are the same for all types of particles, \eg quarks and gluons, up to an overall color factor. This is known as quadratic Casimir scaling \cite{Korchemskaya:1992je,Becher:2009cu,Becher:2009qa,Dixon:2009gx,Gardi:2009qi} and was thought to be valid to all orders in perturbation theory \cite{Becher:2009cu,Becher:2009qa}. In fact, this scaling behavior was confirmed up to three loops and is also a result of this work. However, recent studies raised the possibility that it is violated at higher loop orders and a new, generalized scaling behavior was put forward, starting at four loops \cite{Moch:2018wjh,Ruijl:2016pkm,Catani:2019rvy,Becher:2019avh} and confirmed in \cite{Henn:2019swt,vonManteuffel:2020vjv} using different methods. \\

Reconstructing the renormalization factor $\mathbf{Z}_{\ir}$ is most efficiently done by solving \Eq{eq:GIR_def} iteratively order by order in couplings to reconstruct $\mathbf{Z}_{\ir}$ from a given $\mathbf{\Gamma}_{\ir}$, see \eg \cite{Becher:2009cu,Becher:2009qa, Billis:2019evv}.
It is convenient to define
\begin{equation}
    \mathbf{\Gamma}_{\ir}=\sum_{\Sigma n_i = 1}^{\infty} \boldsymbol{\Gamma}(\{ n_i \}) \prod_i \left(\frac{g_i}{4 \pi}\right)^{2 n_i}, \quad {\Gamma}^\prime=\sum_{\Sigma n_i = 1}^{\infty} {\Gamma}^\prime(\{ n_i \}) \prod_i \left(\frac{g_i}{4 \pi}\right)^{2 n_i},
\end{equation}
where the product over $i$ runs over all coupling $g_i$ (not only gauge couplings) appearing in the theory and
\begin{equation}
    \Gamma^{\prime} \equiv \frac{\partial}{\partial \log \mu} \mathbf{\Gamma}_{\ir} = - \sum_{\{G\}} \sum_{i} C_i^G \gamma_{\rm cusp}^G.
\end{equation}
To efficiently organize our results for the cusp anomalous dimension, we further define
\begin{equation} \label{eq:exp_cusp}
    \gamma_{\rm cusp}^G \equiv \sum_{\Sigma n_i = 1}^{\infty} \gamma_{\rm cusp}^G(\{n_i\}) \prod_i \left(\frac{g_i}{4 \pi}\right)^{2n_i}.
\end{equation}
Up to two loops, \ie ${\Sigma n_i = 2}$, we will find that there are no mixed coupling terms and thus $\gamma_{\rm cusp}^G = \sum_{n_G = 1}^{2} \gamma_{\rm cusp}^G(n_G) \left(\frac{g_i}{4 \pi}\right)^{2n_G}$.
It was believed that the same structure is true also at three loops, see \eg \cite{Chiu:2008vv}.
However, our results explicitly show that this is not the case since we do not find such a simple structure in the theory under consideration in this work. For a discussion, see \Sec{sec:res_anom_dims}.

\subsection{(Non-)Renormalization of Conserved Currents} \label{sec:non_renorm}

Consider a transformation of all fields in a theory
\begin{equation}
    \Phi \to \Phi'= \Phi + \epsilon F[\Phi'] + \mathcal{O}(\epsilon^2),
\end{equation}
leaving the action invariant, $S[\Phi']=S[\Phi]$. The reparametrization invariance of the path integral leads to the Ward identity, the quantum-mechanical equivalent of current conservation \cite{Srednicki:2007qs,Peskin:1995ev,sterman_1993,Pokorski:1987ed,Collins:1984xc},
\begin{equation} \label{eq:Ward_id}
    \frac{\partial}{\partial x_\mu}\left\langle T\left\{j^\mu(x)\prod_{i=1}^{m}\Phi_i(x_i)\right\}\right\rangle=(-i)\sum_{k=1}^{m}\delta(x-x_k)\left\langle T\left\{F_k[\Phi(x_k)]\prod_{i\neq k}^{m}\Phi_i(x_i)\right\}\right\rangle.
\end{equation}
The conserved current $j^\mu$ associated to the symmetry is given by 
\begin{equation} \label{eq:noether}
    j^\mu=\frac{\delta \Lag}{\delta (\partial_\mu\Phi_n)}F_n[\Phi]-K^\mu,
\end{equation}
where $K^\mu$ accounts for a possible transformation of the Lagrangian itself, $\Lag\to\Lag+\partial_\mu K^\mu$. 
\Eq{eq:Ward_id} holds for the bare and renormalized Lagrangian in the same way, such that we can write the Ward identity for both cases. 
On the other hand, we can always express bare operators in terms of renormalized ones using divergent $\rm UV$ renormalization constants, $j^\mu_{\rm bare} = Z_{j}^{\rm UV} j^\mu_{\rm ren}$. Thus, starting with \Eq{eq:Ward_id} with bare quantities, one sees that the left-hand side of the equation is equal to itself times the renormalization constant $Z_{j}^{\rm UV}$ from which we conclude that $Z_{j}^{\rm UV}=1$. \\
From this, we could infer that Green's functions with the insertion of a conserved current, and consequently also the current itself, do not require any renormalization on top of the renormalization of fields and couplings\footnote{For a more detailed derivation of the non-renormalization of conserved currents see \cite{sterman_1993,Pokorski:1987ed,Collins:1984xc}.}. The above argument, however, only holds strictly for the divergence of the Green's functions. In fact, conserved currents can mix with a special class of counter-term operators $O_\mu$ whose divergence vanishes identically, $\partial_\mu O^\mu = $, without using the equations of motion. If those were necessary, the divergent counterterm would introduce new, divergent terms to the right-hand side of \Eq{eq:Ward_id}, spoiling its finiteness. \\
In this work, we are mainly interested in the case where the conserved current is the energy-momentum tensor, which is the conserved Noether current of the space-time translations, 
\begin{equation} \label{eq:tmunu_noether}
    T^{\mu\nu}=\frac{\partial\Lag}{\partial(\partial_\mu\Phi_n)}\partial^\nu\Phi_n-g^{\mu\nu}\Lag,
\end{equation}
which can be made symmetric by adding a total derivative term. \\

If we assume gravity to be non-dynamical, only one identically conserved operator exists, associated with the trace of $T^{\mu \nu}$ \cite{Theil:2023osl}. Therefore, we conclude that the symmetric and traceless part of the stress-energy tensor is not renormalized at any loop order. This implies that, by considering the renormalization of the form factor of $T^{\mu \nu}$, we can calculate the universal infrared anomalous dimensions with no need to separate UV and IR contributions, as the former vanish. \\

Universality of \Eq{eq:IR_anom_dim_general} suggests to extract the value for $\gamma_{\rm cusp}$ and $\gamma_{\rm coll}$ from the simplest non-trivial object one can construct, namely the form factors with two external legs. In the rest of the paper, we show how to compute the IR divergent part of the form factor of the stress-energy tensor
\begin{equation} \label{def:Tmunu}
    \langle T^{\mu \nu} \rangle \equiv \langle 0 |T^{\mu \nu}| \Phi(p_{1}) \bar{\Phi}(p_{2}) \rangle = \includegraphics[width=4.2cm,valign=c]{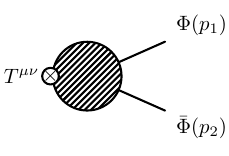},
\end{equation}
for all particle anti-particle pairs $\Phi \bar{\Phi}$ up to three-loop within the theory described in \Sec{sec:theory}. 
Thus, form factors defined via \Eq{def:Tmunu} are three point vertex functions where an external off-shell current with virtuality $s_{12}=(p_1+p_2)^2$ couples to an external particle/anti-particle pair with on-shell momenta $p_1$ and $p_2$.
As explained above, all divergences appearing in this form factor have to be of $\rm IR$ nature \ie
\begin{equation} \label{eq:Tmunu_IR_ren}
    \langle T^{\mu \nu} \rangle_{\rm ren} = \left. \mathbf{Z}_{\rm IR}^{-1} \langle T^{\mu \nu} \rangle_{\rm bare} \right|_{g^{(0)}_i \rightarrow g_i},
\end{equation}
where the bare couplings are expressed in terms of the renormalized ones in order to account for coupling renormalizations.
For the form factor of the energy-momentum tensor $T^{\mu \nu}$ with two external legs with momenta $p_1$ and $p_2$ of pairs of particles and anti-particles \Eq{eq:IR_anom_dim_general} simplifies to
\begin{equation} \label{eq:gir_Tmunu}
\mathbf{\Gamma}_{\ir} = \sum_{\{G\}} C_R^G \gamma_{\rm cusp}^G \log \left( - \frac{s_{12}}{\mu^2} \right) + 2 \gamma^\Phi_{\rm coll}.
\end{equation}
Here we used that the Casimirs of the representations of particle and antiparticle are the same as well as the fact that the collinear anomalous dimension of particle and antiparticle is the same, $\gamma^{\bar{\Phi}}_{\rm coll} = \gamma^\Phi_{\rm coll}$. Thus, we can isolate \textit{all} the cusp and collinear anomalous dimensions of the theory via \Eq{eq:gir_Tmunu} from the renormalization constant $\mathbf{Z}_{\rm IR}$ of \Eq{eq:Tmunu_IR_ren}, which is related to $\mathbf{\Gamma}_{\ir}$ via \Eq{eq:GIR_def}. \\
At this point, we want to emphasize that it is also possible to use other conserved currents such as the gauge current $j^\mu_{\rm B}$, $j^\mu_{\rm W}$ and $j^\mu_{\rm s}$, which also do not $\uv$ renormalize when their respective gauge coupling is switched off, to determine the $\ir$ anomalous dimensions. However, those do not couple to all pairs of external particles we must consider to determine the full $\ir$ structure of a theory such as the SM. The stress-energy tensor $\T$, on the other hand, exists for all pairs of external states, which can be seen as a direct consequence of the fact that gravity couples universally to all types of matter. We, however, still use the gauge currents to cross-check our calculations as described in \Sec{sec:cross_check}.

\section{Model and UV-Renormalization} \label{sec:Model}

\subsection{Lagranian and Action} \label{sec:theory}

We choose to work within a model closely resembling the SM, incorporating all particles of the first-generation SM and a massless Higgs but without right-handed neutrinos. The gauge group is the full $\SU(N_{1})_c\times\SU(N_{2})_W\times\U(1)_Y$\footnote{In this case, we maintain general values for $N_1$ and $N_2$, as this does not increase the complexity of the calculation.} gauge group of the standard model. Our approach can be straightforwardly extended to encompass the full SM.
Therefore, the particle content of our model consists of the gauge bosons, $G$, $W$, and $B$, the Higgs field $H$, a left-handed quark $Q_L$ doublet, and two right-handed quarks $u_R$ and $d_R$, as well as a left-handed lepton doublet $L_L$ and a right-handed lepton $e_R$. The transformation properties of the matter fields are summarized in \Tab{tab:trafos}.
\begin{table}[h] 
\centering
\begin{tabular}{|c|c|c|c|} 
\hline
State & Representation & Representation SM \\
\hline \hline
$Q_L$ & $(\mathbf{N_{1}},\mathbf{N_{2}})_{Y_{QL}}$ &  $(\mathbf{3},\mathbf{2})_{1/6}$ \\
\hline
$d_R$ & $(\mathbf{N_{1}},\mathbf{1})_{Y_{dR}}$ & $(\mathbf{3},\mathbf{1})_{-1/3}$ \\
\hline
$u_R$ & $(\mathbf{N_{1}},\mathbf{1})_{Y_{uR}}$ & $(\mathbf{3},\mathbf{1})_{2/3}$ \\
\hline
$L_L$ & $(\mathbf{1},\mathbf{N_{2}})_{Y_{LL}}$ & $(\mathbf{1},\mathbf{2})_{-1/2}$ \\
\hline
$e_R$ & $(\mathbf{1},\mathbf{1})_{Y_{eR}}$ & $(\mathbf{1},\mathbf{1})_{-1}$ \\
\hline
$H$ & $(\mathbf{1},\mathbf{N_{2}})_{Y_{QL}}$ & $(\mathbf{1},\mathbf{2})_{1/2}$ \\
\hline       
\end{tabular}
\caption{Representations of the matter fields in our model. We denote the representation of the fields under the $\SU(N_{1})_c\times\SU(N_{2})_W\times\U(1)_Y$ group as $(R_{N_1}(\Phi),R_{N_2}(\Phi))_Y$, where the subscript indicates the charge under the Abelian group. In the third column, we quote the explicit values for the physical SM.}
\label{tab:trafos}
\end{table}
We use the well-established $R_\xi$ gauge fixing with $\xi=1$ for both non-abelian gauge groups, also known as Feynman gauge. This introduces ghost fields, $c_G$, and $c_W$, which transform in the adjoint representation under the associated group and as singlets under the others. The Lagrangian for our model, including ghosts, reads
\begin{equation} \label{eq:lag_toy_SM}
\begin{aligned}
    \Lag_M=&-\frac{1}{4}G^{A\mu\nu}G^{A}_{\mu\nu}-\frac{1}{4}W^{I\mu\nu}W^{I}_{\mu\nu}-\frac{1}{4}B^{\mu\nu}B_{\mu\nu}\\
    &-\frac{1}{2}(\partial_\mu G^{A\mu})^2 -\frac{1}{2}(\partial_\mu W^{I\mu})^2 + (\partial_\mu \bar{c}_G)(D^{\mu}c_G)+ (\partial_\mu \bar{c}_W)(D^{\mu}c_W)\\
    &+(D_\mu H)^\dagger(D^\mu H)-\sum_{f}i\bar{f}\slashed{D}f\\
    &+(y_d \bar{Q}_Ld_RH+y_u \bar{Q}_Lu_R\tilde{H}+y_\ell \bar{L}_Le_RH+h.c.)-\frac{\lambda}{2}|H|^4,
\end{aligned}
\end{equation}
where the sum runs over $f=\{Q_L,d_R,u_R,L_L,e_R\}$ and contractions of fundamental group indices are left implicit. The covariant derivative is defined by
\begin{equation}
    D_\mu \phi = \partial_\mu\phi - i g_s G^A_\mu T^A_s \phi - i g_w W^I_\mu T^I_W \phi - i g_1 Y B_\mu \phi,
\end{equation}
for a general field $\phi$ in the fundamental representation of the $\SU(N)$'s that has hypercharge $Y$. We use $g_s$, $g_w$ and $g_1$ as names for the gauge couplings of the $\SU(N_{1})_c$, $\SU(N_{2})_W$ and $\U(1)_Y$ groups, respectively. If a field is in the singlet representation of any of the gauge groups, the respective term in the covariant derivative vanishes. The conjugate Higgs doublet is defined via
\begin{equation}
    \tilde{H}=i\sigma^2H^\ast.
\end{equation}
Note that due to the explicit appearance the $\sigma^2$ matrix, all interactions containing the conjugate Higgs field, \ie the up-type Yukawa coupling, only make sense if we set $N_2=2$ explicitly. Hence, when showing any results, $N_2=2$ has been chosen for all terms containing the coupling $y_u$, while we keep the gauge groups general for all other couplings. To model the effect of multiple fermion generations, we introduce a factor of $N_g$ whenever we encounter a closed fermion loop. \\

We can obtain an explicit expression for the stress-energy tensor $\T$ by performing a weak field expansion of the matter action
\begin{equation}
    S_M = \int d^d x \sqrt{-g} \mathcal{L}_M,
\end{equation}
by separating the metric according to $g_{\mu \nu}=\eta_{\mu \nu}+\kappa h_{\mu \nu}$.
At leading order in $h_{\mu \nu}$ we obtain
\begin{equation} \label{eq:lag_int}
    S_{\rm int} = - \frac{\kappa}{2} \int d^dx h_{\mu \nu} T^{\mu \nu},
\end{equation}
yielding an explicit expression for the symmetriced version $\T$ defined in \Eq{eq:tmunu_noether}\footnote{Extracting $\T$ in this way is equivalent to calculating $T^{\mu\nu}=\frac{-2}{\sqrt{-g}}\frac{\delta S}{\delta g_{\mu\nu}}$}. From the interaction defined in \Eq{eq:lag_int} it is clear how to obtain the form factor of the conserved current $\langle\T\rangle$ at a given loop order in complete analogy to a gauge current with an interaction of the form $j_\mu A^\mu$. \\

As is always the case when considering higher loop renormalization, we also need the UV renormalization, or equivalently, the beta functions of all the involved couplings. Since all tree-level objects we consider do not depend on any coupling\footnote{In the case of the gauge currents, we treat the corresponding gauge coupling as "external" \ie, it does not appear in any diagram apart from the current insertion itself.}, we need the beta functions up to the two-loop order, which we discuss in the next section.

\subsection{UV-Renormalization} \label{sec:UV_renorm}

Here, we will briefly explain the method used to efficiently compute the coupling renormalizations necessary for the calculations in the rest of the paper. 

The coupling renormalization is encoded in the respective beta function
\begin{equation}
    \beta_{i}=\mu^2\frac{dg_i}{d\mu^2},
\end{equation}
where $g_i$ is any renormalized coupling, which in turn is related to the respective bare coupling through the renormalization constant
\begin{equation}
    g_i=Z_i^{-1} g_i^{(0)}.
\end{equation}
The renormalization constants can be uniquely determined such that all correlation functions - together with the renormalization of the external states - are rendered finite. Therefore, a straightforward approach to compute the coupling renormalization would be simply computing all the relevant $n$-point functions and canceling all appearing divergences. Since we are working in a massless theory, not only the UV divergences relevant for the coupling renormalization will appear in the calculation, but also IR divergences, which we have to separate from each other.
Many ways exist to deal with these IR divergences. We decide to introduce an auxiliary mass parameter, serving as an IR cutoff \cite{Misiak:1995,Chetyrkin:1997fm, Zoller:2014}. The idea is to decompose any loop propagator using the identity
\begin{equation}
    \frac{1}{(q+p)^2}=\frac{1}{q^2-M^2}+\frac{-p^2-2p\cdot q-M^2}{q^2-M^2}\frac{1}{(q+p)^2},
\end{equation}
where $p$ and $q$ refer to any combination of loop and external momenta, respectively, and $M$ is an arbitrary, auxiliary mass parameter.
We want to stress here that the second term gives rise to less divergent terms in the UV than the first one due to the higher power of loop momenta in the denominator. Further, the last term contains the original propagator, such that the decomposition can be applied recursively, generating fewer and fewer divergent terms in the process. After each iteration, ignoring the last term, the remaining expression is exactly the Taylor expansion of the original propagator in the UV limit after imbuing it with the additional mass parameter $M$. Then, to efficiently extract the UV divergences, we iteratively decompose every loop propagator in the amplitude until the generated terms are no longer divergent. 
Since $M$ is completely arbitrary, the final result has to be independent of it, such that we can set $M\to0$ in all numerators. This, however, introduces the need to renormalize $M$ itself, especially at higher loop orders, such that all subdivergences are taken care of. The final ingredient for computing the coupling renormalization is to realize that all the remaining propagators do not contain any external momenta in the denominator. Hence, they correspond to pure vacuum bubble diagrams with all massive lines of the same mass $M$. Then, by performing the usual tensor reduction to remove loop momenta from the numerator, up to two loops, all amplitudes can be written in terms of two classes of loop integrals given in \cite{Chetyrkin:1997fm}, which are known and straightforward to evaluate. \\

By employing this procedure for all diagrams, including counterterm diagrams, all non-local subdivergences are canceled, leaving only local UV divergences, which can be canceled by choosing the renormalization appropriately. At the $n$-loop order, the highest divergence is of order $\epsilon^{-n}$. It can be shown that terms with double poles or higher are all related to lower loop divergences, while the beta function can be directly read off the single pole term
\begin{equation}
    \beta_i=\epsilon g_i\sum_n 2^{n}\delta^{(n)}_1,
\end{equation}
where $\delta^{(n)}_1$ denotes the single pole term of the renormalization constant at the $n$-loop order.

As mentioned above, since we set $M\to0$ in the propagator decomposition, we also need to renormalize the mass parameter $M$ for the different fields to be able to cancel all subdivergences at the two-loop level. We find the following one counterterms for $M^2$, where $\epsilon$ is defined via $d=4-2\epsilon$,
\begin{equation}
\begin{gathered}
    \delta_{M^2}^{(B)}=\left(\frac{g_1}{4\pi}\right)^2\frac{1}{3\epsilon}\left[C_{A,2}Y_H^2-4N_g(Y_{eR}^2+C_{A,2}Y_{LL}^2+C_{A,1}(Y_{dR}^2+Y_{uR}^2+C_{A,2}Y_{QL}^2))\right],\\[2ex]
    \delta_{M_2}^{(W)}=\left(\frac{g_w}{4\pi}\right)^2\frac{1}{6\epsilon}\left[1+16C_{A,2}+4(C_{A,1}+1)N_g\right],\\[2ex]
    \delta_{M_2}^{(G)}=-\left(\frac{g_s}{4\pi}\right)^2\frac{2}{3\epsilon}\left[4C_{A,1}+4(C_{A,2}+2)N_g\right],\\[2ex]
    \delta_{M_2}^{(H)}=-\frac{3}{\epsilon}\left[\frac{C_{F,2}g_w^2}{8\pi^2}+\frac{Y_H^2g_1^2}{8\pi^2}-\frac{3C_{A,1}N_g}{8\pi^2}(y_d^2+y_u^2)-\frac{3N_gy_\ell^2}{8\pi^2}+\frac{(C_{A,2}+1)\lambda}{24\pi^2}\right],
\end{gathered}
\end{equation}
corresponding to the Lagrangian terms
\begin{equation}
    \delta\Lag_{M^2}=\frac{M^2}{2}\delta_{M^2}^{(B)}B_\mu B^\mu+\frac{M^2}{2}\delta_{M^2}^{(W)}W^{I}_\mu W^\mu_{I}+\frac{M^2}{2}\delta_{M^2}^{(G)}G^{A}_\mu G^\mu_{A}+\frac{M^2}{2}\delta_{M^2}^{(H)}|H|^2.
\end{equation}
The appearing couplings are defined in \eqref{eq:lag_toy_SM}, $Y_i$ are the $\rm U(1)$ hypercharges and $C_{A,i}$ and $C_{F,i}$ are the Casimirs of the respective gauge groups.
Note that a mass term correction must only be performed for the bosons in the theory. For the gauge bosons, these terms violate the gauge symmetries. However, this is irrelevant, as they only need to cancel unphysical subdivergences. 

\section{Multi-Loop Calculation}\label{sec:computation}

Having set up the model, we are now going to calculate the cusp and collinear anomalous dimensions of the theory up to three loops. As outlined in \Sec{sec:non_renorm}, we do that using the form factor of the stress-energy tenor $\langle T^{\mu \nu} \rangle = \langle 0 |T^{\mu \nu}| \Phi(p_{1}) \bar{\Phi}(p_{2}) \rangle$ with a particle anti-particle pair as external states. In this section, we are going to present the calculation and its technical aspects with focus on the three-loop part of the calculation. 
The $\ir$ anomalous dimensions we directly extract from these form factors via \Eq{eq:Tmunu_IR_ren} and \Eq{eq:gir_Tmunu} are given \App{app:gammas}. In section \App{app:cusp} we show the expressions for the cusp anomalous dimensions for each gauge group up to three-loop, while in \App{app:coll} we present the collinear anomalous dimensions for each particle in the theory up to two loops.

\subsection{Tensor Projection} \label{sec:projection}

As performing higher-order loop calculations can be quite cumbersome, it turns out to be useful to project onto parts of the loop amplitudes that are of interest. This is sometimes referred to as projection technique. In our case, the object of interest, $\langle T^{\mu \nu} \rangle$, has two spacetime indices and up to four group indices in the fundamental or adjoint representation of the $\SU(N)$ gauge groups, depending on the pair of external particles. In the case of particle anti-particle pairs, the group structure will always be proportional to up to two Kroneker deltas $\delta^{A B}$ with the indices $A$ and $B$ in the respective representation, so we can multiply with another delta contracting both indices to get rid of all group indices potentially appearing in our form factor. 

The remaining spacetime indices can be eliminated by multiplying with an appropriate projector. 
This is the same technique we already employed in \cite{Baratella:2022nog}.
We start by writing down all of these structures, which are allowed by the properties of $\T$ (conserved and symmetric) for each combination of external legs. As $\langle \T \rangle$ with two legs emerges from the kinetic terms in the Lagrangian, the only possible combinations of particles that can occur are those where both legs are the same particle. Thus, we just have to distinguish 3 different cases, external scalars, fermions, and vectors, when writing down the general structures. Additionally, we know that $\langle \T \rangle$ resulting from the kinetic term is traceless except for the non-conformally coupled scalar. \\

We split the energy-momentum tensor with two external scalars $S^{\mu \nu} \equiv \langle 0 |T^{\mu \nu}| \phi(p_{1}) \phi(p_{2}) \rangle \allowbreak$ into a part with a vanishing and one with a non-vanishing trace and find the most general structure to be given by
\begin{equation}
    S^{\mu \nu}=T_{S}(s_{12})~T^{\mu\nu}_{S} +T_{S,0}(s_{12})~T^{\mu\nu}_{S,0} ,
\end{equation}
where we defined
\begin{subequations}
    \begin{align}
        T^{\mu\nu}_{S} & =p_1^{\mu}p_2^{\nu}- g^{\mu\nu}~\frac{s_{12}}{4}+\frac{d-2}{2(d-1)}\left(g^{\mu\nu}~\frac{s_{12}}{2}-p_1^{\mu}p_2^{\nu}-p_1^{\mu}p_1^{\nu}\right)+(p_1\leftrightarrow p_2),\\
        T^{\mu\nu}_{S,0} & =-\frac{d-2}{2(d-1)}\left(g^{\mu\nu}~\frac{s_{12}}{2}-p_1^{\mu}p_2^{\nu}-p_1^{\mu}p_1^{\nu}\right)+(p_1\leftrightarrow p_2).
    \end{align}
\end{subequations}
One can show that the $T^{\mu\nu}_{S,0}$ part is exactly generated by the trace of the energy-momentum tensor, whereas $T^{\mu\nu}_{S}$ is exactly the traceless part. We find the projectors fulfilling
\begin{subequations}
\begin{align}
    T_{S}(s_{12}) & = P_{\mu\nu}^{(S)}~S^{\mu\nu}, \\
    T_{S,0}(s_{12}) & = P_{\mu\nu}^{(S,0)}~S^{\mu\nu},
\end{align}
\end{subequations}
to be given by
\begin{subequations}
\begin{align}
    P_{\mu\nu}^{(S)} & = \frac{4 (d-1)}{(d-2)s_{12}^2}~T_{\mu\nu}^{S} ,\\
    P_{\mu\nu}^{(S,0)} & = \frac{4 (d-1)}{(d-2)^2s_{12}^2}~T_{\mu\nu}^{S,0} .
\end{align}
\end{subequations}
The projectors fulfill the relations 
\begin{equation}
     P_{\mu\nu}^{(S)}  P^{\mu\nu}_{(S,0)} = 0, \quad P_{\mu\nu}^{(S)}  \T_S = 1, \quad P_{\mu\nu}^{(S,0)}  \T_{S,0} = 1.
\end{equation}

For fermions, the energy-momentum tensor $F^{\mu \nu} \equiv \langle 0 |T^{\mu \nu}| \psi_{L/R}(p_{1}) \bar{\psi}_{L/R}(p_{2}) \rangle \allowbreak$ has the structure
\begin{equation} \label{eq:Tmunu_F_proj_form}
\begin{aligned}
    {F}^{\mu\nu} & = \bar{v}(p_2) \mathcal{F}^{\mu\nu} P_{L/R} u(p_1) \\
    & = T_F(s_{12})\left[(p_1-p_2)^\mu \bar{v}(p_2) \gamma^{\nu} P_{L/R} u(p_1)+(\mu\leftrightarrow \nu)\right].
\end{aligned}
\end{equation}
To isolate the function $T_F(s_{12})$, we multiply \Eq{eq:Tmunu_F_proj_form} by the conjugate spinors, take the spin sum and trace over fermion indices such that we find
\begin{equation}
    T_F(s_{12})=\frac{4}{(d-2)s_{12}^2}(p_1-p_2)^{\mu}~\text{Tr}\left[\gamma^\nu\slashed{p}_2 \mathcal{F}_{\mu\nu}P_{L/R}\slashed{p}_1\right].
\end{equation}
The prefactor is chosen such that the dependence on the number of spacetime dimensions from the fermion trace is exactly canceled. In this case, it is not possible to write down a projector $P^{(F)}$ with only spacetime indices as in the case with scalar legs. \\

For external vector gauge bosons with $V^{\mu \nu} \equiv \langle 0 |T^{\mu \nu}| V(p_{1}) \bar{V}(p_{2}) \rangle \allowbreak$, the relevant structure appearing is given by
\begin{equation}
\begin{aligned}
    {V}_{\mu\nu}&=\epsilon^\alpha(p_1)\epsilon^\beta(p_2)\mathcal{V}_{\mu\nu\alpha\beta} = T_{V}(s_{12})\epsilon^\alpha(p_1)\epsilon^\beta(p_2) T^V_{\mu\nu\alpha\beta},
\end{aligned}
\end{equation}
where we defined
\begin{equation}
\begin{aligned}
    T^V_{\mu\nu\alpha\beta} = & \frac{s_{12}}{2}\left(g_{\mu\alpha}g_{\nu\beta}+g_{\mu\beta}g_{\nu\alpha}-g_{\mu\nu}g_{\alpha\beta}\right) + p_{1\nu}(p_{2\mu} g_{\alpha\beta}-p_{2\alpha} g_{\mu\beta}) \\
    & +p_{1\mu}(p_{2\nu} g_{\alpha\beta}-p_{2\alpha} g_{\nu\beta}) + p_{1\beta}(p_{2\alpha} g_{\mu\nu}-p_{2\nu} g_{\mu\alpha}-p_{2\mu} g_{\alpha\nu}).
\end{aligned}
\end{equation}
As for the cases above, it is also possible here to find a way of projecting out the relevant function $T_{V}(s_{12})$. We find
\begin{equation}
    T_{V}(s_{12})=P^{(V)}_{\mu\nu\alpha\beta} \mathcal{V}^{\mu\nu\alpha\beta},
\end{equation}
where we defined
\begin{equation}
    P^{(V)}_{\mu\nu\alpha\beta} = \frac{4}{3d^2-14d+16} T^V_{\mu\nu\alpha\beta},
\end{equation}
which is normalized to satisfy
\begin{equation}
    P^{(V)}_{\mu\nu\alpha\beta} T_V^{\mu\nu\alpha\beta} = 1.
\end{equation}

Having projected away any open indices of the form factor independent of the loop order, we are now ready to discuss the details of the loop calculation. At any loop order we use \texttt{Feynrules} \cite{Alloul:2013bka} to implement our model into Mathematica, \texttt{FeynArts} \cite{Hahn:2000kx} and \texttt{FeynCalc} \cite{{Shtabovenko:2023idz,Shtabovenko:2020gxv,Shtabovenko:2016sxi,Mertig:1990an}} to draw and organize the diagrams.

\subsection{One- and Two-Loop Calculation}

At one loop, it is well-known that any integral can be decomposed in a set of scalar basis integrals via the Passarino-Veltman decomposition \cite{Passarino:1978jh}, which we perform using \texttt{FeynCalc}. 
We calculate the necessary basis integrals using Feynman parameters. The same integrals can, for example, also be found in \cite{Ellis:2007qk}. 
To obtain the finite remainder at three loops,
we need the expressions for the one-loop form factor up to $\mathcal{O}(\varepsilon^4)$. \\

\begin{figure}[t]
\begin{center}
\begin{overpic}[height=1.6cm]{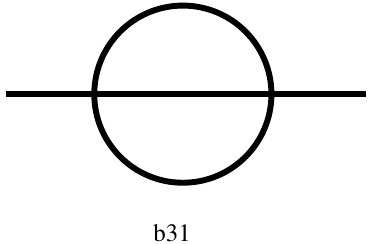}
  \put(30,-1){\colorbox{white}{\small $B_{3,1}$}}
\end{overpic}
\hspace{1cm}
\begin{overpic}[height=1.6cm]{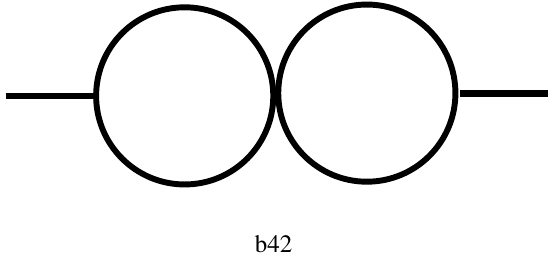}
  \put(35,-1){\colorbox{white}{\small $B_{4,2}$}}
\end{overpic}
\hspace{1cm}
\begin{overpic}[height=1.6cm]{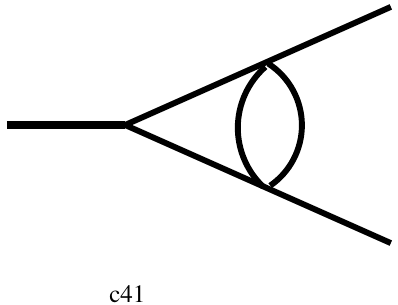}
  \put(25,-1){\colorbox{white}{\small $C_{4,1}$}}
\end{overpic}
\hspace{1cm}
\begin{overpic}[height=1.6cm]{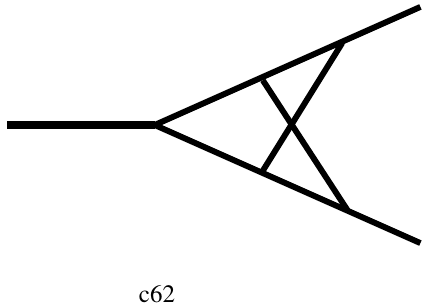}
  \put(30,-1){\colorbox{white}{\small $C_{6,2}$}}
\end{overpic}
\end{center}
\caption{Two-loop master integrals appearing in our the calculation of $\langle \T \rangle$. Diagrams taken from \cite{Gehrmann:2010ue} with permission of the respective authors.}
\label{fig:masters2}
\end{figure}

To simplify the $\SU(N)$ algebras as well as Dirac structures from two loops on, we use \texttt{FormTracer} \cite{Cyrol:2016zqb} as it is computationally more efficient than \texttt{FeynCalc}.
Starting from two-loop on, we cannot rely on Passarino-Veltman decomposition anymore.
After projecting the energy-momentum tensor, we are left with only scalar products of external and loop momenta in the numerators of the loop integrals. These can be further reduced to scalar integrals.
The resulting scalar integrals are, however, not linear independent and can be related to a basis of scalar integrals by using Integration-By-Parts (IBP) \cite{Laporta:2000dsw,Chetyrkin:1981qh} and Lorentz-Invariance (LI) identities. In practice, we use the software \texttt{Kira} \cite{Maierhofer:2017gsa}, which fully automizes the reduction of scalar integrals to the basis integrals. We find that all integrals that appear can be written in terms of four basis integrals. These are exactly the integrals $B_{3,1}$, $B_{4,2}$, $C_{4,1}$, $C_{6,2}$ in \cite{Gehrmann:2010ue} shown in \Fig{fig:masters2}. We do not write down the expressions for the masters here but refer to App. A of \cite{Gehrmann:2010ue}, where they are given explicitly. By the same argument as before at one-loop, we here need the expressions up to $\mathcal{O}(\varepsilon^2)$.

\subsection{Three-Loop Calculation}

For the three-loop calculation, the strategy is the same as at two loops. However, the complexity and computational effort increase, which is why we go into a bit more detail here.
During the calculation of the energy-momentum tensor matrix element, all appearing diagrams can be clustered into different topologies, which
can be mapped to subsets of three auxiliary 12-propagator topologies by simple linear shifts of the loop momenta. These three auxiliary topologies are shown in the first three columns of \Tab{tab:topos}. However, we found it convenient for two 
topologies not to be mapped to any of the auxiliary ones. Instead, we leave them unchanged, apart from completing the integral family by adding independent propagators. These two are shown in the last two columns of \Tab{tab:topos}. To find the mappings between the original and auxiliary topologies, which can be represented in terms of linear shifts of loop momenta, we used both \texttt{Kira} as well as \texttt{Feyncalc}.

\begin{table}[h] 
\centering
\caption{Auxiliary topologies we mapped all but two of the appearing three loop topologies to, as well as the two unchanged topologies. All momenta are understood to be squared in the propagators and $Q=p_1+p_2$.}
\begin{tabular}{lllll} 
AuxTopo 1 & AuxTopo 2 & AuxTopo 3 & Unmapped 1 & Unmapped 2 \\
$k_1$ & $k_1$ & $k_1$ & $k_1$ & $k_1$\\
$k_2$ & $k_2$ & $k_2$ & $k_2$ & $k_2$\\
$k_3$ & $k_3$ & $k_3$ & $k_3$ & $k_3$\\
$k_1-k_2$ & $k_1-k_2$ & $k_1-k_2$ & $k_1+k_2$ & $k_1-k_2$\\
$k_1-k_3$ & $k_1-k_3$ & $k_1-k_3$ & $k_2+p_1$ & $k_2+k_3$\\
$k_2-k_3$ & $k_2-k_3$ & $k_1-k_2-k_3$ & $k_1-Q$ & $k_3+p_1$\\
$k_1-p_1$ & $k_1-k_3-p_2$ & $k_1-p_1$ & $k_1+k_2+k_3-p_2$ & $k_1-Q$\\
$k_1-Q$ & $k_1-Q$ & $k_1-Q$ & $k_1+k_3$ & $k_1-k_2-k_3-Q$\\
$k_2-p_1$ & $k_2-p_1$ & $k_2-p_1$ & $k_1+p_1$ & $k_1+k_3$\\
$k_2-Q$ & $k_1-k_2-p_2$ & $k_2-Q$ & $k_2+k_3$ & $k_1+p_1$\\
$k_3-p_1$ & $k_3-p_1$ & $k_3-p_1$ & $k_2+p_2$ & $k_2+p_1$\\
$k_3-Q$ & $k_3-Q$ & $k_3-Q$ & $k_3+p_1$ & $k_2+p_2$\\
\end{tabular} \label{tab:topos}
\end{table}

After projecting the tensor structure of the form factor $\langle \T \rangle$ with the projectors defined in \Sec{sec:projection} turning all appearing tensor integrals into scalar ones, and after applying symmetry relations, we are left with approximately 123,000 scalar three-loop integrals to solve.
We reduced all Feynman integrals with \texttt{Kira} and found 22 scalar master integrals.

\begin{figure}[H]
\begin{center}
\begin{overpic}[height=1.4cm]{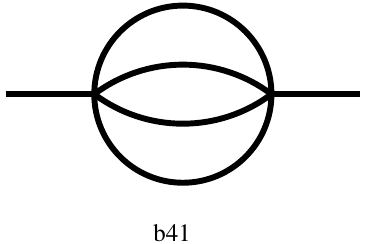}
  \put(30,-1){\colorbox{white}{\small $B_{4,1}$}}
\end{overpic}
\hspace{1cm}
\begin{overpic}[height=1.4cm]{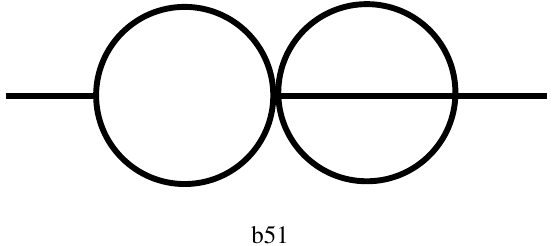}
  \put(35,-1){\colorbox{white}{\small $B_{5,1}$}}
\end{overpic}
\hspace{1cm}
\begin{overpic}[height=1.4cm]{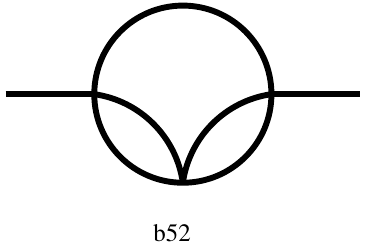}
  \put(30,-1){\colorbox{white}{\small $B_{5,2}$}}
\end{overpic}
\hspace{1cm}
\begin{overpic}[height=1.4cm]{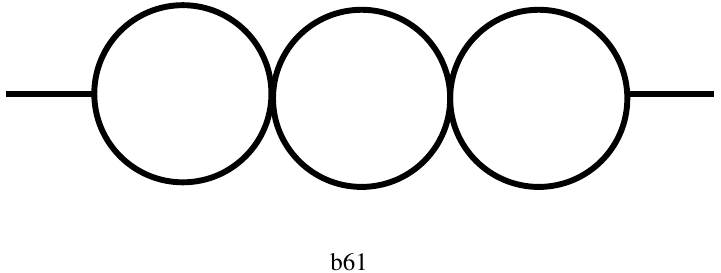}
  \put(40,-1){\colorbox{white}{\small $B_{6,1}$}}
\end{overpic}
\end{center}
\vspace{0.5cm}

\begin{center}
\begin{overpic}[height=1.5cm]{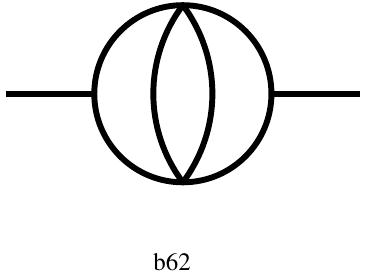}
  \put(30,-1){\colorbox{white}{\small $B_{6,2}$}}
\end{overpic}
\hspace{1cm}
\begin{overpic}[height=1.5cm]{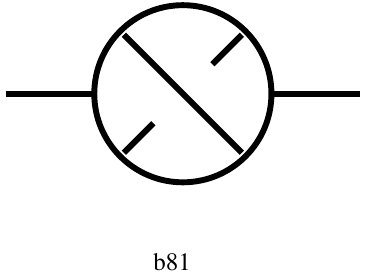}
  \put(30,-1){\colorbox{white}{\small $B_{8,1}$}}
\end{overpic}
\hspace{1cm}
\begin{overpic}[height=1.5cm]{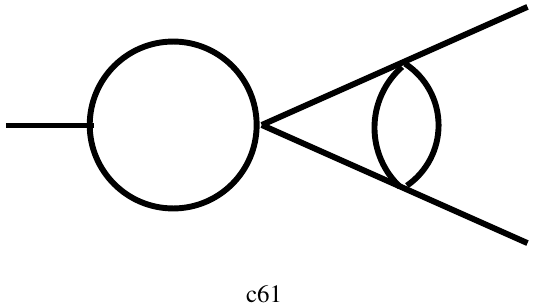}
  \put(30,-1){\colorbox{white}{\small $C_{6,1}$}}
\end{overpic}
\hspace{1cm}
\begin{overpic}[height=1.5cm]{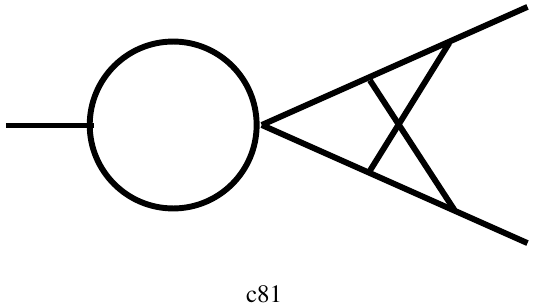}
  \put(30,-1){\colorbox{white}{\small $C_{8,1}$}}
\end{overpic}
\end{center}
\vspace{0.5cm}

\begin{center}
\begin{overpic}[height=2cm]{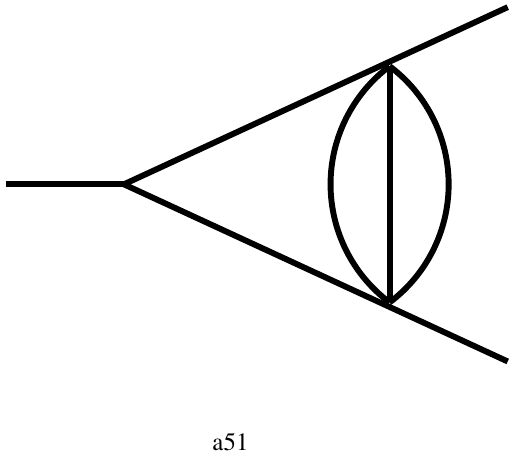}
  \put(30,-1){\colorbox{white}{\small $A_{5,1}$}}
\end{overpic}
\hspace{1cm}
\begin{overpic}[height=2cm]{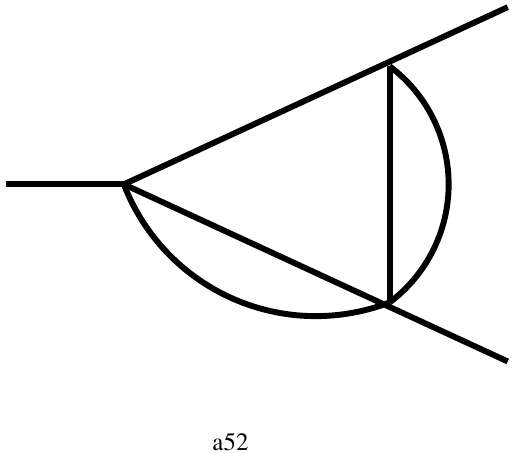}
  \put(30,-1){\colorbox{white}{\small $A_{5,2}$}}
\end{overpic}
\hspace{1cm}
\begin{overpic}[height=2cm]{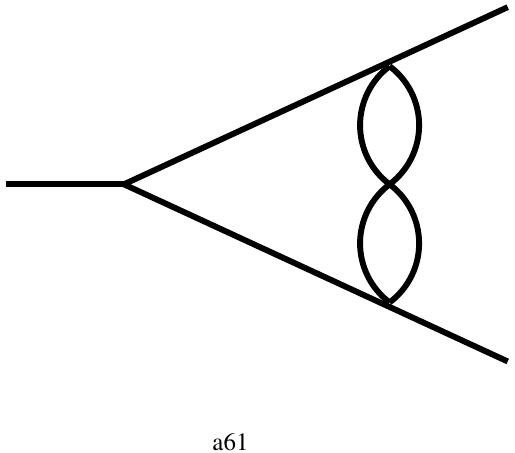}
  \put(30,-1){\colorbox{white}{\small $A_{6,1}$}}
\end{overpic}
\hspace{1cm}
\begin{overpic}[height=2cm]{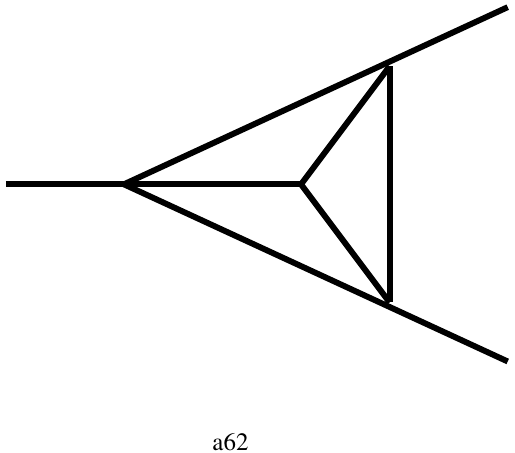}
  \put(30,-1){\colorbox{white}{\small $A_{6,2}$}}
\end{overpic}
\end{center}
\vspace{0.5cm}

\begin{center}
\begin{overpic}[height=2cm]{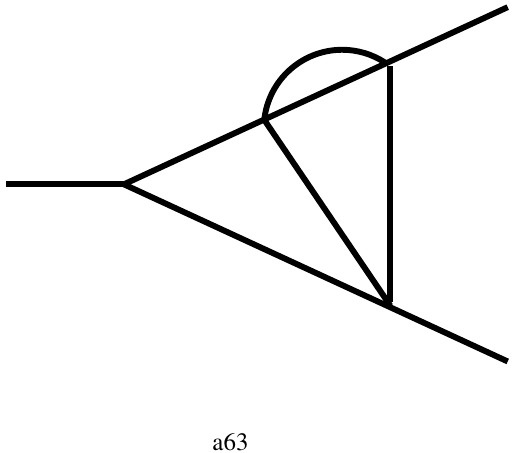}
  \put(30,-1){\colorbox{white}{\small $A_{6,3}$}}
\end{overpic}
\hspace{1cm}
\begin{overpic}[height=2cm]{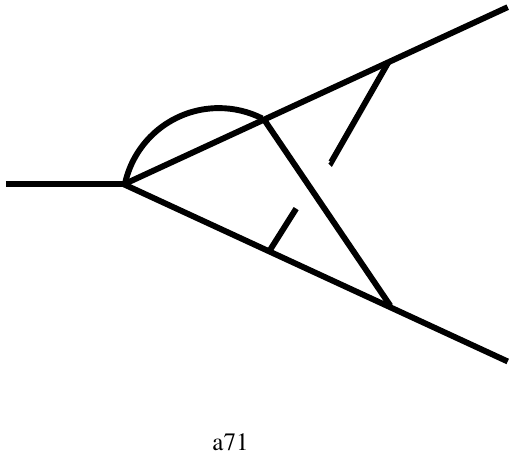}
  \put(30,-1){\colorbox{white}{\small $A_{7,1}$}}
\end{overpic}
\hspace{1cm}
\begin{overpic}[height=2cm]{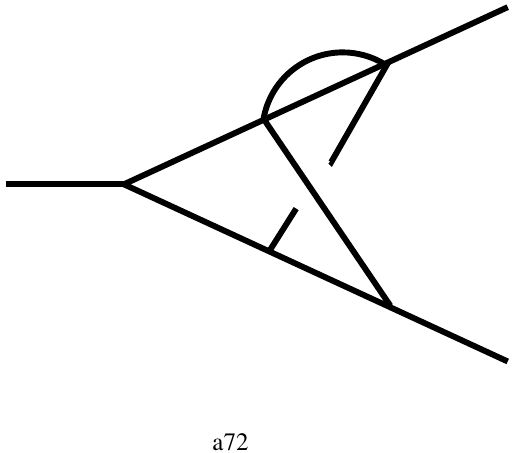}
  \put(30,-1){\colorbox{white}{\small $A_{7,2}$}}
\end{overpic}
\hspace{1cm}
\begin{overpic}[height=2cm]{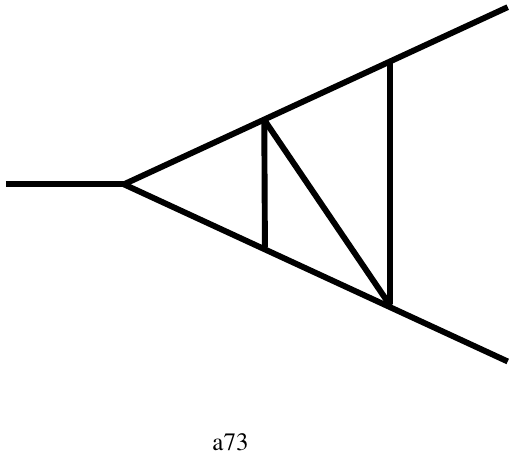}
  \put(30,-1){\colorbox{white}{\small $A_{7,3}$}}
\end{overpic}
\end{center}
\vspace{0.5cm}

\begin{center}
\begin{overpic}[height=2cm]{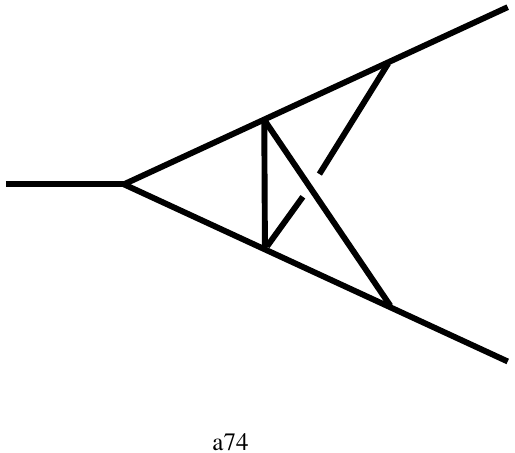}
  \put(30,-1){\colorbox{white}{\small $A_{7,4}$}}
\end{overpic}
\hspace{1cm}
\begin{overpic}[height=2cm]{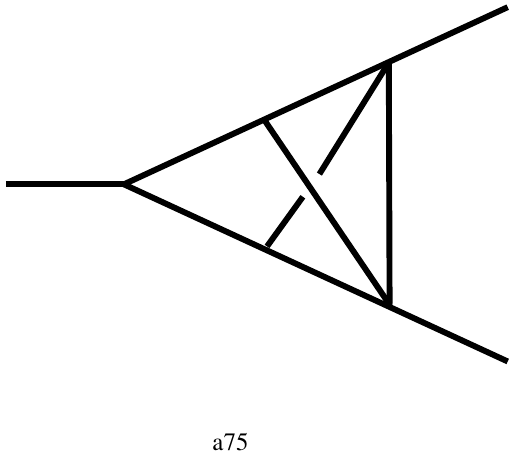}
  \put(30,-1){\colorbox{white}{\small $A_{7,5}$}}
\end{overpic}
\hspace{1cm}
\begin{overpic}[height=2cm]{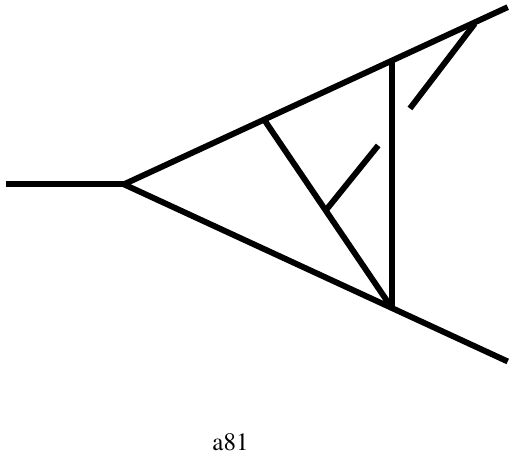}
  \put(30,-1){\colorbox{white}{\small $A_{8,1}$}}
\end{overpic}
\hspace{1cm}
\begin{overpic}[height=2cm]{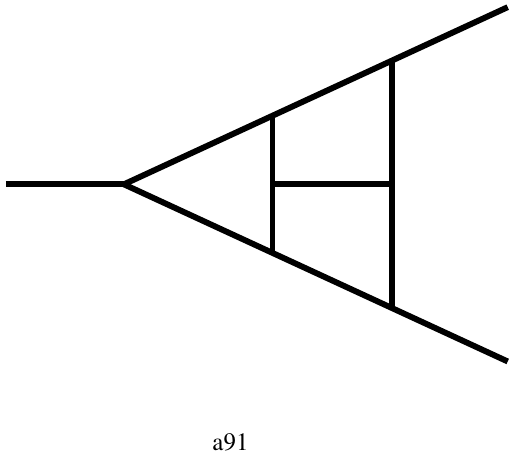}
  \put(30,-1){\colorbox{white}{\small $A_{9,1}$}}
\end{overpic}
\end{center}
\vspace{0.5cm}

\begin{center}
\begin{overpic}[height=2cm]{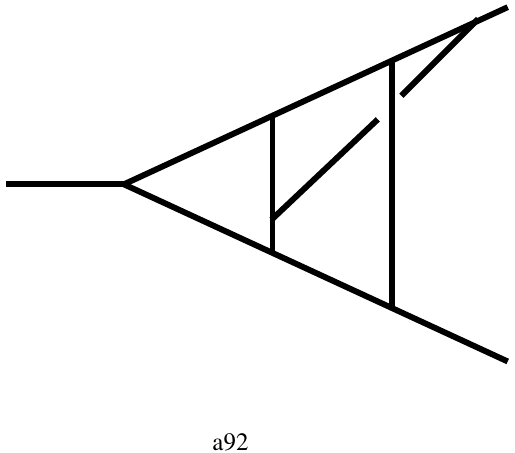}
  \put(30,-1){\colorbox{white}{\small $A_{9,2}$}}
\end{overpic}
\hspace{1cm}
\begin{overpic}[height=2cm]{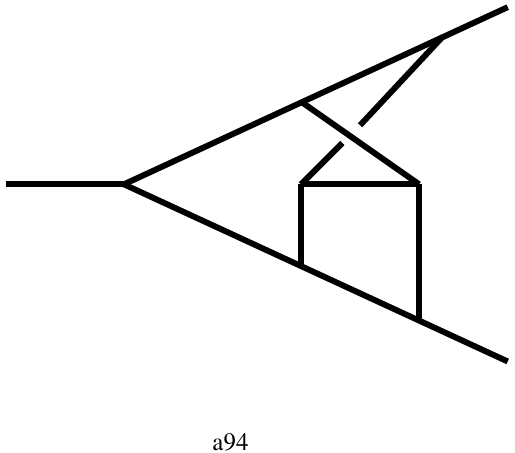}
  \put(30,-1){\colorbox{white}{\small $A_{9,4}$}}
\end{overpic}
\end{center}
\caption{Three-loop master integrals appearing in our the calculation of $\langle \T \rangle$. Diagrams taken from \cite{Gehrmann:2010ue} with permission of the respective authors.}
\label{fig:masters3}
\end{figure}

\newpage

The diagrams corresponding to the master integrals are shown in \cite{Gehrmann:2010ue}, and we collected them in \Fig{fig:masters3}. The analytical expression up to the necessary order in $\varepsilon$ can be found explicitly in \cite{Gehrmann:2010ue}. \\
To wrap up the section, we summarize the numbers of diagrams appearing in the calculation of $\langle \T \rangle$ at each loop order for each particle/anti-particle combination of external legs in \Tab{tab:diags}.

\begin{table}[h] 
\centering
\caption{Number of diagrams for each considered final state at each loop order.}
\begin{tabular}{ccccc} 
 & Tree level & One loop & Two loops & Three loops \\\hline\hline
$Q_L$ & 1 & 10 & 654 & 42,490 \\
$d_R$ & 1 & 6 & 333 & 19,015 \\
$u_R$ & 1 & 6 & 333 & 19,015 \\
$L_L$ & 1 & 6 & 301 & 16,495 \\
$e_R$ & 1 & 4 & 165 & 8,291 \\
$H$   & 1 & 13 & 721 & 47,391 \\
$G$   & 1 & 10 & 511 & 32,242 \\
$W$   & 1 & 11 & 372 & 19,651 \\
$B$   & 1 & 13 & 589 & 36,040 \\\hline\hline
Total & 9 & 79   & 3,979 & 240,630 
\end{tabular}
\label{tab:diags}
\end{table}

\section{Results} \label{sec:results}

\subsection{IR Anomalous Dimensions} \label{sec:res_anom_dims}

Using the form factor $\langle \T \rangle$ calculated in the previous section, we can extract the IR anomalous dimensions. From the absence of divergences on the RHS of \Eq{eq:Tmunu_IR_ren}, we can directly extract $\ZIR$ for each combination of external states. Using \Eq{eq:GIR_def}, we obtain $\GIR$ expanding in powers of couplings. From \Eq{eq:gir_Tmunu}, we can then immediately isolate the cusp and collinear anomalous dimensions. 
We compute for the first time the three-loop cusp anomalous dimension in the presence of a charged massless scalar, including mixed contributions that were previously thought to be vanishing, as well as all the full collinear anomalous dimensions of all particles in the first generation Standard Model.

The collinear anomalous dimensions for each type of particle $\Phi$ in the theory is directly extracted from $\GIR$ corresponding to the respective form factor $\langle T^{\mu \nu} \rangle = \langle 0 |T^{\mu \nu}| \Phi(p_{1}) \bar{\Phi}(p_{2}) \rangle$. The results for $\gamma^\Phi_{\rm coll}$ can be found in \App{app:coll}.
The cusp anomalous dimensions for a given group $G$ can be extracted from $\langle T^{\mu \nu} \rangle$ if the external particle $\Phi$ is charged under any representation of $G$. The resulting $\gamma^G_{\rm cusp}$ have the structure shown in \Eq{eq:exp_cusp}, and we collect them in \App{app:cusp}, where we sort them according to the powers of couplings appearing as in \Eq{eq:exp_cusp}. \\

We find that mixed contributions to the cusp anomalous dimension that have been claimed to be vanishing \cite{Chiu:2008vv} exist and are proportional to the fourth power of any of the gauge couplings, \ie, any term in the cusp anomalous dimension contains at most two different couplings up to three loops, see \App{app:3lcusp}.\footnote{We thank Lance Dixon for a clarifying discussion regarding that point.} \\

Regarding the structure of these mixed terms, we found that they can be treated as a coupling expansion of the pure gauge coupling cusp anomalous dimension in the following sense:
If the final state is charged under a gauge group $G$ with coupling $g$, the corresponding IR anomalous dimension will contain a single-logarithmic term proportional to $g^{2L}$, with $L\leq3$ the loop order under consideration. However, we find that there are also mixed terms proportional to $g^4$. This means that, \eg, the IR anomalous dimension for the quarks contains mixed cusp terms proportional $g_s^4$, while the lepton IR anomalous dimension does not. Further, notice that not all possible combinations of couplings actually appear in the cusp anomalous dimension. We have seen that the "main" coupling is given by the one appearing as a fourth power. Then, the only couplings that can be mixed with the main coupling are such that the fields interacting via this "secondary" coupling are also charged under the gauge group corresponding to the "main" coupling. This means that the strong coupling $g_s$ will not mix with the lepton Yukawa coupling because none of the fields interacting via the lepton Yukawa (leptons, Higgs) are charged under the strong coupling. On the other hand, the hypercharge coupling $g_1$ will mix with all couplings because all matter fields are charged under $\U(1)$. Note that the external state does not necessarily need to interact via the "secondary" coupling at all, as the external state only sets the "main" coupling. As a consequence, the IR anomalous dimension of the $e_R$ state contains a contribution proportional to, \eg $g_1^4y_u^2$, because it is charged under $\U(1)$ while being completely unrelated to the up-type Yukawa coupling.

\begin{figure}
    \centering
    \includegraphics[width=0.45\linewidth]{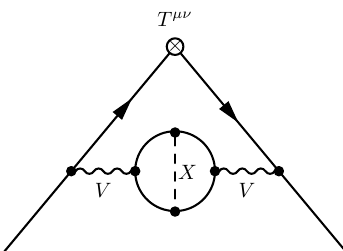}
    \caption{Example diagram contributing to mixed contributions to the cusp anomalous dimension.}
    \label{fig:MixedDiag}
\end{figure}

All of the above-mentioned patterns can be explained by realizing that these types of mixed contributions arise from diagrams where the particle corresponding to the "secondary" interaction is inserted into a subloop, as \eg shown in \Fig{fig:MixedDiag}. Here $X$ can be any of the bosons of our theory, while $V$ corresponds to any of the gauge bosons. If the species of $X$ and $V$ coincide, say all of them are gluons, this diagram would contribute to the generation of the term proportional to $N_{\rm F,\suc}C_{F,1}$ appearing in the pure three loop cusp anomalous dimension. If, however, the species of $X$ and $V$ do not coincide, this diagram would contribute to a corresponding mixed cusp anomalous dimension. In fact, if $X$ is one of the other two gauge bosons not equal to $V$, the corresponding mixed anomalous dimension can straightforwardly be read off from the term containing both the number of fermion generations as well as the fundamental Casimir operators, as can be easily seen in \App{app:gammas}. 
If instead, $X$ corresponds to a scalar, mixed terms containing the Yukawa couplings are generated, which, however, cannot be reconstructed from the pure cusp anomalous dimensions, as there are no such terms for non-gauge couplings.\\
Further, instead of the internal fermion loop, there is also the possibility of a Higgs loop if $V$ corresponds to any of the electroweak bosons. In this case, the corresponding contribution to the mixed cusp anomalous dimension are terms proportional to only the fundamental Casimir and not the number of fermion generations. 

Following this logic, we expect more mixed terms to appear at four loops or higher. In particular, by adding a second type $X$ boson to the loop, we expect to generate mixed terms of the form $g_i^4g_j^4$. Interestingly enough, for these terms, the coupling dependence is not enough to distinguish between the "main" and "secondary" coupling, and instead, the group factors have to be used. Further, if we would add another boson that is neither the same as the original $X$ nor as $V$, we would expect terms of the form $g_i^4g_j^2g_k^2$ to arise. Finally, we also expect mixed terms containing the scalar quartic to appear for the first time at four loops, as this requires adding two more loop lines and, therefore, increases the loop order by two from the starting two-loop diagram.\\

To summarize this discussion as a formula, we recall the expression of the IR anomalous dimension for the case of a particle/anti-particle final state \Eq{eq:gir_Tmunu}. Then, according to our findings, we write the cusp anomalous dimension as
\begin{equation}
    \gamma_{\rm cusp}^G=c^{\tiny(1)}g^2+g^4\left(c^{\tiny(2)}+\sum_{\{g_j\}}d^{j}g_j^2\right)+c^{\tiny(3)}g^6+ \ldots,
 \end{equation}
where the coefficients $c,d$ are just real numbers. Here, we see that the coefficient of the $g^4$ term itself is an expansion in couplings, where the couplings appearing in ${g_j}$ follow the constraint explained above.

Finally, let us note that the above also explains why mixed terms appear at three-loop order for the first time. We explained above that mixed cusp anomalous dimensions arise from two-loop diagrams, which receive perturbative corrections by adding more lines to internal loops. For mixed contributions to arise at two loops, the same would need to happen, but with a one-loop diagram serving as a seed diagram. However, being of one-loop order, no internal loop exists that does not include the external line. Hence, no additional fields in the above sense can be added, and no mixed contributions can arise at two-loop order. 

To conclude this section, we want to mention that it would be highly interesting to perform a study of a model with multiple couplings at four-loop order in order to verify or disprove our expectations for higher-order mixed terms, but this enormous computational task is left for future work.

\subsection{Cross-Checks} \label{sec:cross_check}

We perform multiple cross-checks with the existing literature: We checked the QCD collinear and cusp anomalous dimensions for quarks and gluons up to three loops \cite{Moch:2005id,Becher:2009qa,Becher:2006mr,Gehrmann:2010ue}, as well as the two-loop cusp anomalous dimension in the presence of a charged scalar \cite{Beneke:2019vhz}. 
Contributions that originate only from the gauge interactions of the gauge bosons and fermions should be the same for both the $\SU(N_1)_c$ and $\SU(N_2)_W$ groups, up to trivially replacing group factors. We do indeed explicitly find this analogy in our results.
The QED sector can be checked with \cite{Billis:2019evv}. However, for the case of mixed anomalous dimensions, we disagree with their result as they assumed, based on \cite{Chiu:2008vv}, that there are no mixed-coupling terms in the cusp anomalous dimensions, see discussion in \Sec{sec:res_anom_dims}\footnote{While finalizing this work, Ref.~\cite{AH:2019pyp} has been brought to our attention, where the mixed cusp anomalous dimensions for the $\SU(N_1)_c \times U(1)$ sector have been calculated. Our results for this subsector of our theory are fully consistent with their findings.}. The two-loop beta functions of \Sec{sec:UV_renorm} can be compared with \cite{Luo:2002ey,Bednyakov:2012rb} where we find perfect agreement in the Standard Model limit. \\

Besides comparing with existing literature, a big part of the calculation can also be cross-checked by calculating the form factor of the corresponding (if existing) gauge current $j^\mu_{\rm B}$, $j^\mu_{\rm W}$ and $j^\mu_{\rm s}$ for the same external states as for the stress-energy tensor, while simultaneously switching off the gauge coupling to avoid self-renormalization. One can then extract the anomalous dimensions in complete analogy to what we did for $\T$. We carried out the calculation for the three gauge currents for all possible combinations of external states and found perfect agreement. However, let us stress at this point that using the gauge currents cannot reproduce all results we obtained with $\T$, as there are sectors we cannot access when switching off the respective gauge coupling. This is why it is still important to use $\T$ in order to obtain a complete set of cusp and collinear anomalous dimensions with all mixed-coupling terms, which are not accessible with the gauge currents only. Finally, let us note that the cusp anomalous dimensions resulting from two form factors with different external states charged under the same group are equal, serving as an additional internal consistency check.

\subsection{Supplementary Files}

We provide the analytical results for the cusp and collinear anomalous dimensions in electronic form in the supplementary file \texttt{IRAnomDim\_SM\_3Loop.m}. Additionally, we also provide the beta functions of all coupling of the theory calculated in \Sec{sec:UV_renorm} up to two-loop order in the supplementary file \texttt{BetaFunc\_SM\_2Loop.m}. For the usage of these files, we refer to \App{app:usage_suppl}.

\section{Conclusions} \label{sec:conclusion}

In this paper, we calculated the cusp and collinear anomalous dimension of a theory closely resembling the SM described in \Sec{sec:theory}. We used two important facts of the stress-energy tensor to perform our calculations. First, it couples universally to the SM particles, and therefore, there exists a form factor for any combinations of particle/anti-particle pair as external states. Second, it is conserved and thus not UV-renormalized, which enables us to extract the IR divergences and, thereof, the cusp and collinear anomalous dimensions.

We reproduce the findings from the existing literature where available, with the exception of the mixed cusp anomalous dimensions, which have been overlooked by parts of the literature. Large parts of our results, especially at three loops and most terms where multiple couplings appear, are computed here for the first time.
Where possible, we verified our results by performing an analogous calculation for the gauge currents.

Our results are of particular use for precision calculations, where one needs to subtract unphysical infrared divergences in order to make physical predictions. The IR anomalous dimensions calculated in this paper are made accessible as a supplementary file. With those, it is possible to subtract the IR divergences of any $n$-point amplitude within the theory by constructing the appropriate renormalization factor from the cusp and collinear anomalous dimensions. The generalization of our calculation to the full SM, though straightforward, with the only difference that powers of the scalar Yukawa couplings are replaced by various traces of complex Yukawa matrices. However, this is beyond the scope of this paper and is left for future work.

\acknowledgments

We would like to thank Lance Dixon, Matthias König, Andreas Weiler and Martin Beneke for helpful discussion. We also thank Tobias Huber for assisting with finding minor mistakes in our computational code by providing individual intermediate results from the three-loop computation presented in \cite{Gehrmann:2010ue}.
TT wants to thank Max Zoller and Ulrich Haisch for fruitful and constructive discussions using massive tadpoles for UV renormalization and Johann Usovitsch and Philipp Maierhöfer for helping with understanding certain outputs produced by their \texttt{Kira} software.
MS would like to thank Federico Buccioni for discussion and Cesare Mella for discussion and comments on the manuscript.
This work has been supported by the Collaborative Research Center SFB1258, the Munich Institute for Astro-, Particle and BioPhysics (MIAPbP), and by the Excellence Cluster ORIGINS, which is funded by the Deutsche Forschungsgemeinschaft (DFG, German Research Foundation) under Germany´s Excellence Strategy – EXC 2094 – 39078331. The research of MS is partially
supported by the International Max Planck Research School (IMPRS) on “Elementary Particle Physics”.


\appendix

\section{IR Anomalous Dimensions} \label{app:gammas}

In this appendix, we collect the expressions for the cusp and collinear anomalous dimensions up to three loops, computed in this work. As mentioned in \Sec{sec:Model}, we present the results for general $N_1$ and $N_2$ of the strong gauge group $\SU(N_1)_c$ and the weak gauge group $\SU(N_2)_W$ respectively, except for the terms containing the up Yukawa couplings $y_u$, where we have to take $N_2=2$ as explained in the main text.
Before presenting the results, we want to define the weighted number of fermions contributing to the respective gauge cusp anomalous dimension
\small
\begin{equation}
\begin{gathered}
    N_{\rm F,\suc}=\frac{(C_{A,2}+2)}{2}N_g, \hspace{0.5cm} N_{\rm F,\suw}=\frac{(C_{A,1}+1)}{2}N_g,\\[2ex]
    N^i_{\rm F,U(1)}=N_g\sum_{f} R_{N_1}(f) R_{N_2}(f) Y^i_f=N_g\left[Y_{eR}^i+C_{A,2}Y_{LL}^i+C_{A,1}(Y_{dR}^i+Y_{uR}^i+C_{A,2}Y_{QL}^i)\right].
    \end{gathered}
\label{eq:fermionNumbers}
\end{equation}
\normalsize
as they will be used multiple times throughout this appendix.

\subsection{Cusp Anomalous Dimensions} \label{app:cusp}

\subsubsection{One Loop}
\begin{equation}
    \begin{gathered}
        \gamma_{\tiny \rm \suc}^{\tiny(1)}=4\left(\frac{g_s}{4\pi}\right)^2, \hspace{0.5cm} \gamma_{\tiny \rm \suw}^{\tiny(1)}=4\left(\frac{g_w}{4\pi}\right)^2,\\[2ex]
        \gamma_{\tiny \rm U(1)}^{\tiny(1)}=4\left(\frac{g_1}{4\pi}\right)^2,
    \end{gathered}
\end{equation}

\subsubsection{Two Loops}
\begin{equation}
    \begin{gathered}
        \gamma_{\tiny \rm \suc}^{\tiny(2)}=\left(\frac{g_s}{4\pi}\right)^4\left[C_{A,1}\left(\frac{268}{9}-\frac{4\pi^2}{3}\right)-\frac{40N_{\rm F,\suc}}{9}\right], \\[2ex]
        \gamma_{\tiny \rm \suw}^{\tiny(2)}=\left(\frac{g_w}{4\pi}\right)^4\left[C_{A,1}\left(\frac{268}{9}-\frac{4\pi^2}{3}\right)-\frac{40N_{\rm F,\suw}}{9}-\frac{16}{9}\right],\\[2ex]
        \gamma_{\tiny \rm U(1)}^{\tiny(2)}=\left(\frac{g_1}{4\pi}\right)^4\left[-\frac{40N^2_{F,U(1)}}{9}-2C_{A,2}Y_H^2\frac{16}{9}\right],
    \end{gathered}
\end{equation}

\subsubsection{Three Loops} \label{app:3lcusp}
\small
\begin{equation}
    \begin{gathered}
        \gamma_{\tiny \rm \suc}^{\tiny(3)}=\left(\frac{g_s}{4\pi}\right)^6\left[C_{A,1}^2\left(\frac{490}{9}-\frac{536\pi^2}{27}+\frac{44\pi^4}{45}+\frac{88\zeta_3}{3}\right)+N_{\rm F,\suc}C_{A,1}\left(-\frac{836}{27}+\frac{80\pi^2}{27}-\frac{112\zeta_3}{3}\right)\right.\\
        \left.+N_{\rm F,\suc}C_{F,1}\left(-\frac{110}{3}+32\zeta_3\right)-\frac{16N_{\rm F,\suc}^2}{27}\right], \\[2ex]
        \gamma_{\tiny \rm \suw}^{\tiny(3)}=\left(\frac{g_w}{4\pi}\right)^6\left[C_{A,2}^2\left(\frac{490}{9}-\frac{536\pi^2}{27}+\frac{44\pi^4}{45}+\frac{88\zeta_3}{3}\right)+N_{\rm F,\suw}C_{A,2}\left(-\frac{836}{27}+\frac{80\pi^2}{27}-\frac{112\zeta_3}{3}\right)\right.\\
        \left.+N_{\rm F,\suw}C_{F,2}\left(-\frac{110}{3}+32\zeta_3\right)-\frac{16N_{\rm F,\suw}^2}{27}\right.\\
        \left.+C_{A,2}\left(-\frac{265}{27}+\frac{32\pi^2}{27}-\frac{28\zeta_3}{3}\right)+C_{F,2}\left(-\frac{86}{3}+8\zeta_3\right)+\frac{4}{27}\right],\\[2ex]
        \gamma_{\tiny \rm U(1)}^{\tiny(3)}=\left(\frac{g_1}{4\pi}\right)^6\left[\frac{16 C_{A,2}^2Y_H^4}{27}+C_{A,2}Y_H^4\left(-\frac{172}{3}+16\zeta_3\right)+N^4_{\rm F,U(1)}\left(-\frac{110}{3}+32\zeta_3\right)-\frac{16\left(N^2_{\rm F,U(1)}\right)^2}{27}\right],
    \end{gathered}
\end{equation}
\normalsize

\hspace{1cm}

\begin{equation}
    \begin{gathered}
        \gamma_{\tiny \rm \suc/U(1)}^{\tiny(3)}=\left(\frac{g_s}{4\pi}\right)^4\left(\frac{g_1}{4\pi}\right)^2\left[N_g\left(Y_{dR}^2+Y_{uR}^2+C_{A,2}Y_{QL}^2\right)\left(-\frac{55}{3}+16\zeta_3\right)\right],\\[2ex]
        \gamma_{\tiny \rm \suc/\suw}^{\tiny(3)}=\left(\frac{g_s}{4\pi}\right)^4\left(\frac{g_w}{4\pi}\right)^2\left[C_{A,2}C_{F,2}N_g\left(-\frac{55}{3}+16\zeta_3\right)\right]\\[2ex]
        \gamma_{\tiny \rm \suc/yd}^{\tiny(3)}=11C_{A,2}N_g\left(\frac{g_s}{4\pi}\right)^4\left(\frac{y_d}{4\pi}\right)^2, \hspace{0.5cm} \gamma_{\tiny \rm \suc/yu}^{\tiny(3)}=22N_g\left(\frac{g_s}{4\pi}\right)^4\left(\frac{y_u}{4\pi}\right)^2,
    \end{gathered}
\end{equation}

\begin{equation}
    \begin{gathered}
        \gamma_{\tiny \rm \suw/U(1)}^{\tiny(3)}=\left(\frac{g_w}{4\pi}\right)^4\left(\frac{g_1}{4\pi}\right)^2\left[Y_H^2\left(-\frac{86}{3}+8\zeta_3\right)+N_g\left(Y_{LL}^2+C_{A,1}Y_{QL}^2\right)\left(-\frac{55}{3}+16\zeta_3\right)\right],\\[2ex]
        \gamma_{\tiny \rm \suw/\suc}^{\tiny(3)}=\left(\frac{g_w}{4\pi}\right)^4\left(\frac{g_s}{4\pi}\right)^2\left[C_{A,1}C_{F,1}N_g\left(-\frac{55}{3}+16\zeta_3\right)\right]\\[2ex]
        \gamma_{\tiny \rm \suw/yd}^{\tiny(3)}=\frac{17C_{A,1}N_g}{2}\left(\frac{g_w}{4\pi}\right)^4\left(\frac{y_d}{4\pi}\right)^2, \hspace{0.5cm} \gamma_{\tiny \rm \suw/yu}^{\tiny(3)}=\frac{17C_{A,1}N_g}{2}\left(\frac{g_w}{4\pi}\right)^4\left(\frac{y_u}{4\pi}\right)^2, \\[2ex]
        \gamma_{\tiny \rm \suw/y\ell}^{\tiny(3)}=\frac{17}{2}\left(\frac{g_w}{4\pi}\right)^4\left(\frac{y_\ell}{4\pi}\right)^2,
    \end{gathered}
\end{equation}

\begin{equation}
    \begin{gathered}
        \gamma_{\tiny \rm U(1)/\suc}^{\tiny(3)}=\left(\frac{g_1}{4\pi}\right)^4\left(\frac{g_s}{4\pi}\right)^2\left[C_{A,1}C_{F,1}N_g\left(Y_{dR}^2+Y_{uR}^2+C_{A,2}Y_{QL}^2\right)\left(-\frac{110}{3}+32\zeta_3\right)\right],\\[2ex]
        \gamma_{\tiny \rm U(1)/\suw}^{\tiny(3)}=\left(\frac{g_1}{4\pi}\right)^4\left(\frac{g_w}{4\pi}\right)^2C_{A,2}C_{F,2}\left[Y_H^2\left(-\frac{172}{3}+16\zeta_3\right)\right.\\[2ex]
        \left.+N_g\left(Y_{LL}^2+C_{A,1}Y_{QL}^2\right)\left(-\frac{110}{3}+32\zeta_3\right)\right],\\[2ex]
        \gamma_{\tiny \rm U(1)/yd}^{\tiny(3)}=\left(\frac{g_1}{4\pi}\right)^4\left(\frac{y_d}{4\pi}\right)^2\left[C_{A,1}C_{A,2}N_g\left(6Y_H^2+11(Y_{dR}^2+Y_{QL}^2)\right)\right],\\[2ex]
        \gamma_{\tiny \rm U(1)/yu}^{\tiny(3)}=\left(\frac{g_1}{4\pi}\right)^4\left(\frac{y_u}{4\pi}\right)^2\left[2C_{A,1}N_g\left(6Y_H^2+11(Y_{uR}^2+Y_{QL}^2)\right)\right],\\[2ex]
        \gamma_{\tiny \rm U(1)/y\ell}^{\tiny(3)}=\left(\frac{g_1}{4\pi}\right)^4\left(\frac{y_\ell}{4\pi}\right)^2\left[C_{A,2}N_g\left(6Y_H^2+11(Y_{eR}^2+Y_{LL}^2)\right)\right].
    \end{gathered}
\end{equation}

\subsection{Collinear Anomalous Dimensions} \label{app:coll}

\subsubsection{One Loop}
\begin{equation}
    \gamma^{\tiny{(1)}}_{\tiny{G}}=\left(\frac{g_s}{4\pi}\right)^2\left[-\frac{11C_{A,1}}{3}+\frac{N_{\rm F,\suc}}{3}\right]
\end{equation}
\begin{equation}
    \gamma^{\tiny{(1)}}_{\tiny{W}}=\left(\frac{g_w}{4\pi}\right)^2\left[-\frac{11C_{A,2}}{3}+\frac{N_{\rm F,\suw}}{3}+\frac{1}{6}\right]
\end{equation}
\begin{equation}
    \gamma^{\tiny{(1)}}_{\tiny{B}}=\left(\frac{g_1}{4\pi}\right)^2\left[\frac{2N_{\rm F,U(1)}^2}{3}+\frac{C_{A,2}Y_H^2}{3}\right]
\end{equation}
\begin{equation}
    \gamma^{\tiny{(1)}}_{\tiny{u_R}}=-3C_{F,1}\left(\frac{g_s}{4\pi}\right)^2-3Y_{uR}^2\left(\frac{g_1}{4\pi}\right)^2+\left(\frac{y_u}{4\pi}\right)^2
\end{equation}
\begin{equation}
    \gamma^{\tiny{(1)}}_{\tiny{d_R}}=-3C_{F,1}\left(\frac{g_s}{4\pi}\right)^2-3Y_{dR}^2\left(\frac{g_1}{4\pi}\right)^2+\frac{C_{A,2}}{2}\left(\frac{y_d}{4\pi}\right)^2
\end{equation}
\begin{equation}
    \gamma^{\tiny{(1)}}_{\tiny{Q_L}}=-3C_{F,1}\left(\frac{g_s}{4\pi}\right)^2-3C_{F,2}\left(\frac{g_w}{4\pi}\right)^2-3Y_{QL}^2\left(\frac{g_1}{4\pi}\right)^2+\left(\frac{y_u}{4\pi}\right)^2+\frac{C_{A,2}}{2}\left(\frac{y_d}{4\pi}\right)^2
\end{equation}
\begin{equation}
    \gamma^{\tiny{(1)}}_{\tiny{e_R}}=-3Y_{eR}^2\left(\frac{g_1}{4\pi}\right)^2+\frac{C_{A,2}}{2}\left(\frac{y_\ell}{4\pi}\right)^2+\left(\frac{y_d}{4\pi}\right)^2
\end{equation}
\begin{equation}
    \gamma^{\tiny{(1)}}_{\tiny{Q_L}}=-3C_{F,2}\left(\frac{g_w}{4\pi}\right)^2-3Y_{LL}^2\left(\frac{g_1}{4\pi}\right)^2+\frac{C_{A,2}}{2}\left(\frac{y_\ell}{4\pi}\right)^2
\end{equation}
\begin{equation}
    \gamma^{\tiny{(1)}}_{\tiny{H}}=-4C_{F,2}\left(\frac{g_w}{4\pi}\right)^2-4Y_{H}^2\left(\frac{g_1}{4\pi}\right)^2+C_{A,1}N_g\left(\frac{y_u}{4\pi}\right)^2+C_{A,1}N_g\left(\frac{y_d}{4\pi}\right)^2+N_g\left(\frac{y_\ell}{4\pi}\right)^2
\end{equation}

\subsubsection{Two Loops}
\begin{equation}
    \begin{gathered}
        \gamma^{\tiny{(2)}}_{\tiny{G}}=\left(\frac{g_s}{4\pi}\right)^4\left[C_{A,1}^2\left(-\frac{692}{27}+\frac{11\pi^2}{18}+2\zeta_3\right)+C_{A,1}N_{\rm F,\suc}\left(\frac{128}{27}-\frac{\pi^2}{9}\right)+2C_{F,1}N_{\rm F,\suc}\right]\\[2ex]
        +\left(\frac{g_s}{4\pi}\right)^2\left(\frac{g_1}{4\pi}\right)^2\left[N_g\left(Y_{dR}^2+Y_{uR}^2+C_{A,1}Y_{QL}^2\right)\right]+C_{A,2}C_{F,2}N_g\left(\frac{g_s}{4\pi}\right)^2\left(\frac{g_w}{4\pi}\right)^2\\[2ex]
        -C_{A,2}N_g\left(\frac{g_s}{4\pi}\right)^2\left(\frac{y_d}{4\pi}\right)^2-2N_g\left(\frac{g_s}{4\pi}\right)^2\left(\frac{y_u}{4\pi}\right)^2
    \end{gathered}
\end{equation}

\begin{equation}
    \begin{gathered}
        \gamma^{\tiny{(2)}}_{\tiny{W}}=\left(\frac{g_w}{4\pi}\right)^4\left[C_{A,2}^2\left(-\frac{692}{27}+\frac{11\pi^2}{18}+2\zeta_3\right)+C_{A,2}\left(\frac{44}{27}-\frac{\pi^2}{9}\right)+2C_{F,2}\right.\\[2ex]
        \left.+C_{A,2}N_{\rm F,\suw}\left(\frac{128}{27}-\frac{\pi^2}{9}\right)+2C_{F,2}N_{\rm F,\suw}\right]\\[2ex]
        +\left(\frac{g_w}{4\pi}\right)^2\left(\frac{g_1}{4\pi}\right)^2\left[N_g\left(2Y_{H}^2+N_g(Y_{LL}^2+C_{A,1}Y_{QL}^2)\right)\right]\\[2ex]
        +C_{A,1}C_{F,1}N_g\left(\frac{g_w}{4\pi}\right)^2\left(\frac{g_s}{4\pi}\right)^2-\frac{N_g}{2}\left(\frac{g_w}{4\pi}\right)^2\left(\frac{y_\ell}{4\pi}\right)^2\\[2ex]
        -\frac{C_{A,1}N_g}{2}\left(\frac{g_w}{4\pi}\right)^2\left(\frac{y_d}{4\pi}\right)^2-\frac{C_{A,1}N_g}{2}\left(\frac{g_w}{4\pi}\right)^2\left(\frac{y_u}{4\pi}\right)^2
    \end{gathered}
\end{equation}

\begin{equation}
    \begin{gathered}
        \gamma^{\tiny{(2)}}_{\tiny{B}}=\left(\frac{g_1}{4\pi}\right)^4\left[4C_{A,2}Y_H^2+2N_{\rm F,U(1)}^2\right]+\left(\frac{g_1}{4\pi}\right)^2\left(\frac{g_s}{4\pi}\right)^2\left[2C_{A,1}C_{F,1}N_g\left(Y_{dR}^2+Y_{uR}^2+C_{A,2}Y_{QL}^2\right)\right]\\[2ex]
        +\left(\frac{g_1}{4\pi}\right)^2\left(\frac{g_w}{4\pi}\right)^2\left[4C_{A,1}C_{F,1}Y_H^2+2C_{A,1}C_{F,1}N_g\left(Y_{LL}^2+C_{A,1}Y_{LL}^2\right)\right]\\[2ex]
        -\left(\frac{g_1}{4\pi}\right)^2\left(\frac{y_d}{4\pi}\right)^2\left[C_{A,2}C_{A,3}N_g\left(Y_{dR}^2+Y_{QL}^2\right)\right]\\[2ex]
        -\left(\frac{g_1}{4\pi}\right)^2\left(\frac{y_u}{4\pi}\right)^2\left[2C_{A,3}N_g\left(Y_{uR}^2+Y_{QL}^2\right)\right]-\left(\frac{g_1}{4\pi}\right)^2\left(\frac{y_\ell}{4\pi}\right)^2\left[C_{A,3}N_g\left(Y_{eR}^2+Y_{LL}^2\right)\right]
    \end{gathered}
\end{equation}

\begin{equation}
    \begin{gathered}
    \gamma^{\tiny{(2)}}_{\tiny{u_R}}=\left(\frac{g_s}{4\pi}\right)^4\left[C_{F,1}^2\left(-\frac{3}{2}+2\pi^2-24\zeta_3\right)+C_{F,1}C_{A,1}\left(-\frac{961}{54}-\frac{11\pi^2}{6}+26\zeta_3\right)\right.\\[2ex]
    \left.+C_{F,1}N_{\rm F,\suc}\left(\frac{65}{27}+\frac{\pi^2}{3}\right)\right]\\[2ex]
    +\left(\frac{g_1}{4\pi}\right)^4\left[Y_{uR}^4\left(-\frac{3}{2}+2\pi^2-24\zeta_3\right)+Y_{uR}^2N_{\rm F,U(1)}^2\left(\frac{65}{27}+\frac{\pi^2}{3}\right)\right.\\[2ex]
    \left.+Y_{uR}^2Y_H^2C_{A,2}\left(\frac{167}{54}+\frac{\pi^2}{6}\right)\right]\\[2ex]
    +\left(\frac{y_u}{4\pi}\right)^4\left[-\frac{3C_{A,1}N_g}{2}-\frac{1}{4}\right]+\left(\frac{g_s}{4\pi}\right)^2\left(\frac{y_u}{4\pi}\right)^2\left[C_{F,1}\left(-4-\frac{2\pi^2}{3}\right)\right]\\[2ex]
    +\left(\frac{g_1}{4\pi}\right)^2\left(\frac{y_u}{4\pi}\right)^2\left[Y_{uR}^2\left(-\frac{3}{2}-\frac{2\pi^2}{3}\right)+11Y_{H}^2-\frac{5Y_{QL}^2}{2}\right]\\[2ex]
    +\frac{51}{8}\left(\frac{g_w}{4\pi}\right)^2\left(\frac{y_u}{4\pi}\right)^2+\left(\frac{g_s}{4\pi}\right)^2\left(\frac{g_1}{4\pi}\right)^2\left[C_{F,1}Y_{uR}\left(-3+4\pi^2-48\zeta_3\right)\right]\\[2ex]
    +\left(\frac{y_u}{4\pi}\right)^2\left(\frac{y_d}{4\pi}\right)^2\left[-\frac{3C_{A,1}N_g}{2}-\frac{1}{4}\right]-\frac{3N_g}{2}\left(\frac{y_u}{4\pi}\right)^2\left(\frac{y_\ell}{4\pi}\right)^2
    \end{gathered}
\end{equation}
\vspace{7cm}

\begin{equation}
    \begin{gathered}
    \gamma^{\tiny{(2)}}_{\tiny{d_R}}=\left(\frac{g_s}{4\pi}\right)^4\left[C_{F,1}^2\left(-\frac{3}{2}+2\pi^2-24\zeta_3\right)+C_{F,1}C_{A,1}\left(-\frac{961}{54}-\frac{11\pi^2}{6}+26\zeta_3\right)\right.\\[2ex]
    \left.+C_{F,1}N_{\rm F,\suc}\left(\frac{65}{27}+\frac{\pi^2}{3}\right)\right]\\[2ex]
    +\left(\frac{g_1}{4\pi}\right)^4\left[Y_{dR}^4\left(-\frac{3}{2}+2\pi^2-24\zeta_3\right)+Y_{dR}^2N_{\rm F,U(1)}^2\left(\frac{65}{27}+\frac{\pi^2}{3}\right)\right.\\[2ex]
    \left.+Y_{dR}^2Y_H^2C_{A,2}\left(\frac{167}{54}+\frac{\pi^2}{6}\right)\right]\\[2ex]
    +\left(\frac{y_d}{4\pi}\right)^4\left[-\frac{3C_{A,1}C_{A,2}N_g}{4}-\frac{C_{A,2}}{8}\right]+\left(\frac{g_s}{4\pi}\right)^2\left(\frac{y_d}{4\pi}\right)^2\left[C_{F,1}C_{A,2}\left(-2-\frac{\pi^2}{3}\right)\right]\\[2ex]
    +\left(\frac{g_1}{4\pi}\right)^2\left(\frac{y_d}{4\pi}\right)^2\left[Y_{uR}^2C_{A,2}\left(-\frac{3}{4}-\frac{\pi^2}{3}\right)+C_{A,2}\left(\frac{11}{2}Y_{H}^2-\frac{5Y_{QL}^2}{4}\right)\right]\\[2ex]
    +\frac{17C_{A,2}C_{F,2}}{4}\left(\frac{g_w}{4\pi}\right)^2\left(\frac{y_d}{4\pi}\right)^2+\left(\frac{g_s}{4\pi}\right)^2\left(\frac{g_1}{4\pi}\right)^2\left[C_{F,1}Y_{uR}\left(-3+4\pi^2-48\zeta_3\right)\right]\\[2ex]
    +\left(\frac{y_d}{4\pi}\right)^2\left(\frac{y_u}{4\pi}\right)^2\left[-\frac{3C_{A,1}N_g}{2}-\frac{1}{4}\right]-\frac{3C_{A,2}N_g}{4}\left(\frac{y_d}{4\pi}\right)^2\left(\frac{y_\ell}{4\pi}\right)^2
    \end{gathered}
\end{equation}
\vspace{7cm}

\begin{equation}
    \begin{gathered}
    \gamma^{\tiny{(2)}}_{\tiny{Q_L}}=\left(\frac{g_s}{4\pi}\right)^4\left[C_{F,1}^2\left(-\frac{3}{2}+2\pi^2-24\zeta_3\right)+C_{F,1}C_{A,1}\left(-\frac{961}{54}-\frac{11\pi^2}{6}+26\zeta_3\right)\right.\\[2ex]
    \left.+C_{F,1}N_{\rm F,\suc}\left(\frac{65}{27}+\frac{\pi^2}{3}\right)\right]\\[2ex]
    +\left(\frac{g_w}{4\pi}\right)^4\left[C_{F,2}^2\left(-\frac{3}{2}+2\pi^2-24\zeta_3\right)+C_{F,2}C_{A,2}\left(-\frac{961}{54}-\frac{11\pi^2}{6}+26\zeta_3\right)\right.\\[2ex]
    \left.+C_{F,2}N_{\rm F,\suw}\left(\frac{65}{27}+\frac{\pi^2}{3}\right)+C_{F,2}\left(\frac{167}{54}+\frac{\pi^2}{6}\right)\right]\\[2ex]
    +\left(\frac{g_1}{4\pi}\right)^4\left[Y_{QL}^4\left(-\frac{3}{2}+2\pi^2-24\zeta_3\right)+Y_{QL}^2N_{\rm F,U(1)}^2\left(\frac{65}{27}+\frac{\pi^2}{3}\right)\right.\\[2ex]
    \left.+Y_{QL}^2Y_H^2C_{A,2}\left(\frac{167}{54}+\frac{\pi^2}{6}\right)\right]\\[2ex]
    +\left(\frac{y_d}{4\pi}\right)^4\left[-\frac{3C_{A,1}N_g}{4}-\frac{C_{A,2}}{8}\right]+\left(\frac{y_u}{4\pi}\right)^4\left[-\frac{3C_{A,1}N_g}{4}-\frac{1}{4}\right]\\[2ex]
    +\left(\frac{g_s}{4\pi}\right)^2\left(\frac{y_d}{4\pi}\right)^2\left[C_{F,1}\left(-2-\frac{\pi^2}{3}\right)\right]+\left(\frac{g_s}{4\pi}\right)^2\left(\frac{y_u}{4\pi}\right)^2\left[C_{F,1}\left(-2-\frac{\pi^2}{3}\right)\right]\\[2ex]
    +\left(\frac{g_w}{4\pi}\right)^2\left(\frac{y_d}{4\pi}\right)^2\left[C_{F,2}\left(\frac{19}{4}-\frac{\pi^2}{3}\right)\right]+\left(\frac{g_s}{4\pi}\right)^2\left(\frac{y_u}{4\pi}\right)^2\left[\frac{57}{16}-\frac{\pi^2}{4}\right]\\[2ex]
    +\left(\frac{g_1}{4\pi}\right)^2\left(\frac{y_d}{4\pi}\right)^2\left[Y_{QL}^2\left(-\frac{3}{4}-\frac{\pi^2}{3}\right)+\frac{11}{2}Y_{H}^2-\frac{5Y_{dR}^2}{4}\right]\\[2ex]
    +\left(\frac{g_1}{4\pi}\right)^2\left(\frac{y_u}{4\pi}\right)^2\left[Y_{QL}^2\left(-\frac{3}{4}-\frac{\pi^2}{3}\right)+\frac{11}{2}Y_{H}^2-\frac{5Y_{uR}^2}{4}\right]\\[2ex]
    +\left(\frac{g_s}{4\pi}\right)^2\left(\frac{g_w}{4\pi}\right)^2\left[C_{F,1}C_{F,2}\left(-3+4\pi^2-48\zeta_3\right)\right]\\[2ex]
    +\left(\frac{g_s}{4\pi}\right)^2\left(\frac{g_1}{4\pi}\right)^2\left[C_{F,1}Y_{QL}^2\left(-3+4\pi^2-48\zeta_3\right)\right]\\[2ex]
    +\left(\frac{g_w}{4\pi}\right)^2\left(\frac{g_1}{4\pi}\right)^2\left[C_{F,2}Y_{QL}^2\left(-3+4\pi^2-48\zeta_3\right)\right]\\[2ex]
    +\frac{3C_{A,1}N_g}{2}\left(\frac{y_d}{4\pi}\right)^2\left(\frac{y_u}{4\pi}\right)^2+\frac{3N_g}{4}\left(\frac{y_d}{4\pi}\right)^2\left(\frac{y_\ell}{4\pi}\right)^2+\frac{3N_g}{4}\left(\frac{y_u}{4\pi}\right)^2\left(\frac{y_\ell}{4\pi}\right)^2
    \end{gathered}
\end{equation}

\begin{equation}
    \begin{gathered}
        \gamma^{\tiny{(2)}}_{\tiny{e_R}}=\left(\frac{g_1}{4\pi}\right)^4\left[Y_{eR}^4\left(-\frac{3}{2}+2\pi^2-24\zeta_3\right)+Y_{eR}^2N_{\rm F,U(1)}^2\left(\frac{65}{27}+\frac{\pi^2}{3}\right)\right.\\[2ex]
        \left.+Y_{eR}^2Y_H^2C_{A,2}\left(\frac{167}{54}+\frac{\pi^2}{6}\right)\right]\\[2ex]
        +\left(\frac{y_\ell}{4\pi}\right)^4\left[-\frac{3C_{A,1}C_{A,2}N_g}{4}-\frac{C_{A,2}}{8}\right]+\frac{17C_{A,2}C_{F,2}}{4}\left(\frac{g_w}{4\pi}\right)^2\left(\frac{y_\ell}{4\pi}\right)^2\\[2ex]
        +\left(\frac{g_1}{4\pi}\right)^2\left(\frac{y_\ell}{4\pi}\right)^2\left[Y_{eR}^2C_{A,2}\left(-\frac{3}{4}-\frac{\pi^2}{3}\right)+C_{A,2}\left(\frac{11}{2}Y_{H}^2-\frac{5Y_{LL}^2}{4}\right)\right]\\[2ex]
        -\frac{3C_{A,2}N_g}{4}\left(\frac{y_\ell}{4\pi}\right)^2\left(\frac{y_d}{4\pi}\right)^2-\frac{3N_g}{2}\left(\frac{y_\ell}{4\pi}\right)^2\left(\frac{y_u}{4\pi}\right)^2
    \end{gathered}
\end{equation}

\begin{equation}
    \begin{gathered}
    \gamma^{\tiny{(2)}}_{\tiny{L_L}}=\left(\frac{g_w}{4\pi}\right)^4\left[C_{F,2}^2\left(-\frac{3}{2}+2\pi^2-24\zeta_3\right)+C_{F,2}C_{A,2}\left(-\frac{961}{54}-\frac{11\pi^2}{6}+26\zeta_3\right)\right.\\[2ex]
    \left.+C_{F,2}N_{\rm F,\suw}\left(\frac{65}{27}+\frac{\pi^2}{3}\right)+C_{F,2}\left(\frac{167}{54}+\frac{\pi^2}{6}\right)\right]\\[2ex]
    +\left(\frac{g_1}{4\pi}\right)^4\left[Y_{LL}^4\left(-\frac{3}{2}+2\pi^2-24\zeta_3\right)+Y_{LL}^2N_{\rm F,U(1)}^2\left(\frac{65}{27}+\frac{\pi^2}{3}\right)\right.\\[2ex]
    \left.+Y_{LL}^2Y_H^2C_{A,2}\left(\frac{167}{54}+\frac{\pi^2}{6}\right)\right]\\[2ex]
    +\left(\frac{y_\ell}{4\pi}\right)^4\left[-\frac{3N_g}{4}-\frac{C_{A,2}}{8}\right]+\left(\frac{g_w}{4\pi}\right)^2\left(\frac{y_\ell}{4\pi}\right)^2\left[C_{F,2}\left(\frac{19}{4}-\frac{\pi^2}{3}\right)\right]\\[2ex]
    +\left(\frac{g_1}{4\pi}\right)^2\left(\frac{y_\ell}{4\pi}\right)^2\left[Y_{LL}^2\left(-\frac{3}{4}-\frac{\pi^2}{3}\right)+\frac{11}{2}Y_{H}^2-\frac{5Y_{eR}^2}{4}\right]\\[2ex]
    +\left(\frac{g_w}{4\pi}\right)^2\left(\frac{g_1}{4\pi}\right)^2\left[C_{F,2}Y_{LL}^2\left(-3+4\pi^2-48\zeta_3\right)\right]\\[2ex]
    +\frac{3C_{A,1}N_g}{4}\left(\frac{y_\ell}{4\pi}\right)^2\left(\frac{y_d}{4\pi}\right)^2+\frac{3C_{A,1}N_g}{4}\left(\frac{y_\ell}{4\pi}\right)^2\left(\frac{y_u}{4\pi}\right)^2
    \end{gathered}
\end{equation}

\begin{equation}
    \begin{gathered}
    \gamma^{\tiny{(2)}}_{\tiny{H}}=\left(\frac{g_w}{4\pi}\right)^4\left[C_{F,2}^2\left(\frac{7}{2}+\frac{8\pi^2}{3}-24\zeta_3\right)+C_{F,2}C_{A,2}\left(-\frac{3365}{108}-\frac{11\pi^2}{6}+26\zeta_3\right)\right.\\[2ex]
    \left.+C_{F,2}N_{\rm F,\suw}\left(\frac{113}{27}+\frac{\pi^2}{3}\right)+C_{F,2}\left(\frac{223}{108}+\frac{\pi^2}{12}\right)\right]\\[2ex]
    +\left(\frac{g_1}{4\pi}\right)^4\left[Y_{H}^4C_{A,2}\left(\frac{233}{54}+\frac{\pi^2}{6}\right)+Y_{H}^2N_{\rm F,U(1)}^2\left(\frac{113}{27}+\frac{\pi^2}{3}\right)\right.\\[2ex]
    \left.+Y_H^4\left(\frac{7}{2}+\frac{8\pi^2}{3}-24\zeta_3\right)\right]\\[2ex]
    +\left(\frac{y_d}{4\pi}\right)^4\left[-\frac{3C_{A,1}N_g}{4}(1+C_{A,2})\right]+\left(\frac{y_u}{4\pi}\right)^4\left[-\frac{9C_{A,1}N_g}{4}\right]\\[2ex]+\left(\frac{y_\ell}{4\pi}\right)^4\left[-\frac{3N_g}{4}(1+C_{A,2})\right]+\left(\frac{\lambda}{16\pi^2}\right)^2\left[\frac{1}{2}(1+C_{A,2})\right]\\[2ex]
    +\left(\frac{g_s}{4\pi}\right)^2\left(\frac{y_d}{4\pi}\right)^2\left[5C_{A,1}C_{F,1}N_g\right]+\left(\frac{g_s}{4\pi}\right)^2\left(\frac{y_u}{4\pi}\right)^2\left[5C_{A,1}C_{F,1}N_g\right]\\[2ex]
    +\left(\frac{g_w}{4\pi}\right)^2\left(\frac{y_d}{4\pi}\right)^2\left[C_{A,1}C_{F,2}N_g\left(\frac{5}{2}-\frac{2\pi^2}{3}\right)\right]+\left(\frac{g_w}{4\pi}\right)^2\left(\frac{y_u}{4\pi}\right)^2\left[C_{A,1}N_g\left(\frac{15}{8}-\frac{\pi^2}{2}\right)\right]\\[2ex]
    +\left(\frac{g_w}{4\pi}\right)^2\left(\frac{y_\ell}{4\pi}\right)^2\left[C_{F,2}N_g\left(\frac{5}{2}-\frac{2\pi^2}{3}\right)\right]\\[2ex]
    +\left(\frac{g_1}{4\pi}\right)^2\left(\frac{y_d}{4\pi}\right)^2\left[C_{A,1}N_g\left(-\frac{2\pi Y_H^2}{3}+\frac{5Y_{QL}^2}{2}+\frac{5Y_{dR}^2}{2}\right)\right]\\[2ex]
    +\left(\frac{g_1}{4\pi}\right)^2\left(\frac{y_u}{4\pi}\right)^2\left[C_{A,1}N_g\left(-\frac{2\pi Y_H^2}{3}+\frac{5Y_{QL}^2}{2}+\frac{5Y_{uR}^2}{2}\right)\right]\\[2ex]
    +\left(\frac{g_1}{4\pi}\right)^2\left(\frac{y_\ell}{4\pi}\right)^2\left[N_g\left(-\frac{2\pi Y_H^2}{3}+\frac{5Y_{LL}^2}{2}+\frac{5Y_{eR}^2}{2}\right)\right]\\[2ex]
    +\left(\frac{g_w}{4\pi}\right)^2\left(\frac{g_1}{4\pi}\right)^2\left[C_{F,2}Y_H^2\left(7+\frac{16\pi^2}{3}-48\zeta_3\right)\right]\\[2ex]
    +\left(\frac{y_d}{4\pi}\right)^2\left(\frac{y_u}{4\pi}\right)^2\left[\frac{C_{A,1}N_g}{2}\right]
    \end{gathered}
\end{equation}

\subsubsection{Three Loops}

The three-loop collinear anomalous dimensions can be found in the supplementary file \texttt{IRAnomDim\_SM\_3Loop.m}.

\newpage

\section{Usage of the Supplementary Files} \label{app:usage_suppl}

The supplementary files can be downloaded from our  GitHub repository\footnote{\hbox{\url{https://github.com/michael-stadlbauer/AnomalousDimensions.git}}}~\href{https://github.com/michael-stadlbauer/AnomalousDimensions.git}{\faGithub}.
Upon loading the file \texttt{IRAnomDim\_SM\_3Loop.m} into \texttt{Mathematica} using the usual \texttt{Get} command, the two variables \texttt{$\gamma$Cusp} and \texttt{$\gamma$Coll} are created in the workspace. Both of these are functions that accept at least one and up to two arguments. The first specifies the gauge group (for \texttt{$\gamma$Cusp}) or the external particle (for \texttt{$\gamma$Coll}). If no second argument is provided, \eg \texttt{$\gamma$Coll["B"]}, the full anomalous dimension up to three loops is returned, while a certain loop order can be specified using the second argument. Note, using the second argument returns only a single loop order, i.e. \texttt{$\gamma$Coll["B",2]} returns only the two loop collinear anomalous dimension of the $B$ gauge boson, not up to the two loop order. The allowed values for the two types of anomalous dimensions are given in \Tab{tab:allowedArgs}.
\begin{table}[h] 
\centering
\caption{Allowed arguments for the to functions \texttt{$\gamma$Coll} and \texttt{$\gamma$Cusp} in the supplementary file. Note that they need to be passed as strings. Note, that the cusp anomalous dimensions are called with the SM values of the gauge groups \ie $\SU(3)$ and $\SU(2)$, while the results are still given for general $\SU(N_1)_c$ and $\SU(N_2)_W$ as in the rest of this work.}
\scalebox{0.87}{
\begin{tabular}{cc} 
\texttt{$\gamma$Coll} & \texttt{$\gamma$Cusp} \\\hline\hline
     & U1  \\
     & SU2  \\
     & SU3  \\
     & SU3xSU2  \\
B    & SU3xU1  \\
W    & SU3xYd  \\
G    & SU3xYu  \\
eR   & SU2xSU3  \\
LL   & SU2xU1  \\
dR   & SU2xYd  \\
uR   & SU2xYu  \\
QL   & SU2xYl  \\
H    & U1xSU2  \\
     & U1xSU3  \\
     & U1xYd  \\
     & U1xYu  \\
     & U1xYl  \\\hline\hline
\end{tabular}}
\label{tab:allowedArgs}
\end{table}

After loading the file \texttt{BetaFunc\_SM\_2Loop.m} into a \texttt{Mathematica} session, the function \texttt{beta} is created. This function can be called with a single argument specifying the desired coupling. The allowed values are in the set \{"gs", "gw", "g1", "yd", "yu", "yl", "$\lambda$"\}. To return the beta function of the strong coupling, for example, we would call \texttt{beta["gs"]}. Note that always the full beta function up to the two-loop order is returned. This also includes the scalar quartic coupling. In fact, it turns out this coupling appears only from two loops on in the IR anomalous dimensions. Therefore, we would need only its one-loop beta function, but we still included the two-loop results for completeness.

\newpage

\bibliographystyle{JHEP}
\addcontentsline{toc}{section}{Bibliography}
\bibliography{bibliography}

\providecommand{\href}[2]{#2}\begingroup\raggedright\begin{thebibliography}{10}

\bibitem{Bauer:2000yr}
C.~W. Bauer, S.~Fleming, D.~Pirjol, and I.~W. Stewart, {\it {An Effective field
  theory for collinear and soft gluons: Heavy to light decays}},  {\em Phys.
  Rev. D} {\bf 63} (2001) 114020,
  [\href{http://arxiv.org/abs/hep-ph/0011336}{{\tt hep-ph/0011336}}].

\bibitem{Bauer:2001ct}
C.~W. Bauer and I.~W. Stewart, {\it {Invariant operators in collinear effective
  theory}},  {\em Phys. Lett. B} {\bf 516} (2001) 134--142,
  [\href{http://arxiv.org/abs/hep-ph/0107001}{{\tt hep-ph/0107001}}].

\bibitem{Bauer:2001yt}
C.~W. Bauer, D.~Pirjol, and I.~W. Stewart, {\it {Soft collinear factorization
  in effective field theory}},  {\em Phys. Rev. D} {\bf 65} (2002) 054022,
  [\href{http://arxiv.org/abs/hep-ph/0109045}{{\tt hep-ph/0109045}}].

\bibitem{Bauer:2000ew}
C.~W. Bauer, S.~Fleming, and M.~E. Luke, {\it {Summing Sudakov logarithms in $B
  \to X_s \gamma $in effective field theory.}},  {\em Phys. Rev. D} {\bf 63}
  (2000) 014006, [\href{http://arxiv.org/abs/hep-ph/0005275}{{\tt
  hep-ph/0005275}}].

\bibitem{Beneke:2002ph}
M.~Beneke, A.~P. Chapovsky, M.~Diehl, and T.~Feldmann, {\it {Soft collinear
  effective theory and heavy to light currents beyond leading power}},  {\em
  Nucl. Phys. B} {\bf 643} (2002) 431--476,
  [\href{http://arxiv.org/abs/hep-ph/0206152}{{\tt hep-ph/0206152}}].

\bibitem{Beneke:2002ni}
M.~Beneke and T.~Feldmann, {\it {Multipole expanded soft collinear effective
  theory with nonAbelian gauge symmetry}},  {\em Phys. Lett. B} {\bf 553}
  (2003) 267--276, [\href{http://arxiv.org/abs/hep-ph/0211358}{{\tt
  hep-ph/0211358}}].

\bibitem{Matsuura:1988sm}
T.~Matsuura, S.~C. van~der Marck, and W.~L. van Neerven, {\it {The Calculation
  of the Second Order Soft and Virtual Contributions to the Drell-Yan
  Cross-Section}},  {\em Nucl. Phys. B} {\bf 319} (1989) 570--622.

\bibitem{Matsuura:1987wt}
T.~Matsuura and W.~L. van Neerven, {\it {Second Order Logarithmic Corrections
  to the {Drell-Yan} Cross-section}},  {\em Z. Phys. C} {\bf 38} (1988) 623.

\bibitem{Kramer:1986sg}
G.~Kramer and B.~Lampe, {\it {Two Jet Cross-Section in e+ e- Annihilation}},
  {\em Z. Phys. C} {\bf 34} (1987) 497. [Erratum: Z.Phys.C 42, 504 (1989)].

\bibitem{Harlander:2000mg}
R.~V. Harlander, {\it {Virtual corrections to g g ---\ensuremath{>} H to two
  loops in the heavy top limit}},  {\em Phys. Lett. B} {\bf 492} (2000) 74--80,
  [\href{http://arxiv.org/abs/hep-ph/0007289}{{\tt hep-ph/0007289}}].

\bibitem{Gehrmann:2005pd}
T.~Gehrmann, T.~Huber, and D.~Maitre, {\it {Two-loop quark and gluon
  form-factors in dimensional regularisation}},  {\em Phys. Lett. B} {\bf 622}
  (2005) 295--302, [\href{http://arxiv.org/abs/hep-ph/0507061}{{\tt
  hep-ph/0507061}}].

\bibitem{Moch:2005tm}
S.~Moch, J.~A.~M. Vermaseren, and A.~Vogt, {\it {Three-loop results for quark
  and gluon form-factors}},  {\em Phys. Lett. B} {\bf 625} (2005) 245--252,
  [\href{http://arxiv.org/abs/hep-ph/0508055}{{\tt hep-ph/0508055}}].

\bibitem{Gehrmann:2010ue}
T.~Gehrmann, E.~W.~N. Glover, T.~Huber, N.~Ikizlerli, and C.~Studerus, {\it
  {Calculation of the quark and gluon form factors to three loops in QCD}},
  {\em JHEP} {\bf 06} (2010) 094, [\href{http://arxiv.org/abs/1004.3653}{{\tt
  arXiv:1004.3653}}].

\bibitem{vonManteuffel:2020vjv}
A.~von Manteuffel, E.~Panzer, and R.~M. Schabinger, {\it {Cusp and collinear
  anomalous dimensions in four-loop QCD from form factors}},  {\em Phys. Rev.
  Lett.} {\bf 124} (2020), no.~16 162001,
  [\href{http://arxiv.org/abs/2002.04617}{{\tt arXiv:2002.04617}}].

\bibitem{Baratella:2022nog}
P.~Baratella, S.~Maggio, M.~Stadlbauer, and T.~Theil, {\it {Two-loop infrared
  renormalization with on-shell methods}},  {\em Eur. Phys. J. C} {\bf 83}
  (2023), no.~8 751, [\href{http://arxiv.org/abs/2207.08831}{{\tt
  arXiv:2207.08831}}].

\bibitem{deFlorian:2013sza}
D.~de~Florian, M.~Mahakhud, P.~Mathews, J.~Mazzitelli, and V.~Ravindran, {\it
  {Quark and gluon spin-2 form factors to two-loops in QCD}},  {\em JHEP} {\bf
  02} (2014) 035, [\href{http://arxiv.org/abs/1312.6528}{{\tt
  arXiv:1312.6528}}].

\bibitem{Ahmed:2015qia}
T.~Ahmed, G.~Das, P.~Mathews, N.~Rana, and V.~Ravindran, {\it {Spin-2 Form
  Factors at Three Loop in QCD}},  {\em JHEP} {\bf 12} (2015) 084,
  [\href{http://arxiv.org/abs/1508.05043}{{\tt arXiv:1508.05043}}].

\bibitem{Becher:2009cu}
T.~Becher and M.~Neubert, {\it {Infrared singularities of scattering amplitudes
  in perturbative QCD}},  {\em Phys. Rev. Lett.} {\bf 102} (2009) 162001,
  [\href{http://arxiv.org/abs/0901.0722}{{\tt arXiv:0901.0722}}]. [Erratum:
  Phys.Rev.Lett. 111, 199905 (2013)].

\bibitem{Becher:2009qa}
T.~Becher and M.~Neubert, {\it {On the Structure of Infrared Singularities of
  Gauge-Theory Amplitudes}},  {\em JHEP} {\bf 06} (2009) 081,
  [\href{http://arxiv.org/abs/0903.1126}{{\tt arXiv:0903.1126}}]. [Erratum:
  JHEP 11, 024 (2013)].

\bibitem{Catani:1996jh}
S.~Catani and M.~H. Seymour, {\it {The Dipole formalism for the calculation of
  QCD jet cross-sections at next-to-leading order}},  {\em Phys. Lett. B} {\bf
  378} (1996) 287--301, [\href{http://arxiv.org/abs/hep-ph/9602277}{{\tt
  hep-ph/9602277}}].

\bibitem{Catani:1996vz}
S.~Catani and M.~H. Seymour, {\it {A General algorithm for calculating jet
  cross-sections in NLO QCD}},  {\em Nucl. Phys. B} {\bf 485} (1997) 291--419,
  [\href{http://arxiv.org/abs/hep-ph/9605323}{{\tt hep-ph/9605323}}]. [Erratum:
  Nucl.Phys.B 510, 503--504 (1998)].

\bibitem{Catani:1998bh}
S.~Catani, {\it {The Singular behavior of QCD amplitudes at two loop order}},
  {\em Phys. Lett. B} {\bf 427} (1998) 161--171,
  [\href{http://arxiv.org/abs/hep-ph/9802439}{{\tt hep-ph/9802439}}].

\bibitem{Korchemskaya:1992je}
I.~A. Korchemskaya and G.~P. Korchemsky, {\it {On lightlike Wilson loops}},
  {\em Phys. Lett. B} {\bf 287} (1992) 169--175.

\bibitem{Dixon:2009gx}
L.~J. Dixon, {\it {Matter Dependence of the Three-Loop Soft Anomalous Dimension
  Matrix}},  {\em Phys. Rev. D} {\bf 79} (2009) 091501,
  [\href{http://arxiv.org/abs/0901.3414}{{\tt arXiv:0901.3414}}].

\bibitem{Gardi:2009qi}
E.~Gardi and L.~Magnea, {\it {Factorization constraints for soft anomalous
  dimensions in QCD scattering amplitudes}},  {\em JHEP} {\bf 03} (2009) 079,
  [\href{http://arxiv.org/abs/0901.1091}{{\tt arXiv:0901.1091}}].

\bibitem{Moch:2018wjh}
S.~Moch, B.~Ruijl, T.~Ueda, J.~A.~M. Vermaseren, and A.~Vogt, {\it {On quartic
  colour factors in splitting functions and the gluon cusp anomalous
  dimension}},  {\em Phys. Lett. B} {\bf 782} (2018) 627--632,
  [\href{http://arxiv.org/abs/1805.09638}{{\tt arXiv:1805.09638}}].

\bibitem{Ruijl:2016pkm}
B.~Ruijl, T.~Ueda, J.~A.~M. Vermaseren, J.~Davies, and A.~Vogt, {\it {First
  Forcer results on deep-inelastic scattering and related quantities}},  {\em
  PoS} {\bf LL2016} (2016) 071, [\href{http://arxiv.org/abs/1605.08408}{{\tt
  arXiv:1605.08408}}].

\bibitem{Catani:2019rvy}
S.~Catani, D.~De~Florian, and M.~Grazzini, {\it {Soft-gluon effective coupling
  and cusp anomalous dimension}},  {\em Eur. Phys. J. C} {\bf 79} (2019), no.~8
  685, [\href{http://arxiv.org/abs/1904.10365}{{\tt arXiv:1904.10365}}].

\bibitem{Becher:2019avh}
T.~Becher and M.~Neubert, {\it {Infrared singularities of scattering amplitudes
  and N$^{3}$LL resummation for $n$-jet processes}},  {\em JHEP} {\bf 01}
  (2020) 025, [\href{http://arxiv.org/abs/1908.11379}{{\tt arXiv:1908.11379}}].

\bibitem{Henn:2019swt}
J.~M. Henn, G.~P. Korchemsky, and B.~Mistlberger, {\it {The full four-loop cusp
  anomalous dimension in $\mathcal{N}=4$ super Yang-Mills and QCD}},  {\em
  JHEP} {\bf 04} (2020) 018, [\href{http://arxiv.org/abs/1911.10174}{{\tt
  arXiv:1911.10174}}].

\bibitem{Billis:2019evv}
G.~Billis, F.~J. Tackmann, and J.~Talbert, {\it {Higher-Order Sudakov
  Resummation in Coupled Gauge Theories}},  {\em JHEP} {\bf 03} (2020) 182,
  [\href{http://arxiv.org/abs/1907.02971}{{\tt arXiv:1907.02971}}].

\bibitem{Chiu:2008vv}
J.-y. Chiu, R.~Kelley, and A.~V. Manohar, {\it {Electroweak Corrections using
  Effective Field Theory: Applications to the LHC}},  {\em Phys. Rev. D} {\bf
  78} (2008) 073006, [\href{http://arxiv.org/abs/0806.1240}{{\tt
  arXiv:0806.1240}}].

\bibitem{Srednicki:2007qs}
M.~Srednicki, {\em {Quantum field theory}}.
\newblock Cambridge University Press, 1, 2007.

\bibitem{Peskin:1995ev}
M.~E. Peskin and D.~V. Schroeder, {\em {An Introduction to quantum field
  theory}}.
\newblock Addison-Wesley, Reading, USA, 1995.

\bibitem{sterman_1993}
G.~Sterman, {\em An Introduction to Quantum Field Theory}.
\newblock Cambridge University Press, 1993.

\bibitem{Pokorski:1987ed}
S.~Pokorski, {\em {GAUGE FIELD THEORIES}}.
\newblock Cambridge University Press, 1, 2005.

\bibitem{Collins:1984xc}
J.~C. Collins, {\em {Renormalization}: {An Introduction to Renormalization, The
  Renormalization Group, and the Operator Product Expansion}}, vol.~26 of {\em
  Cambridge Monographs on Mathematical Physics}.
\newblock Cambridge University Press, Cambridge, 1986.

\bibitem{Theil:2023osl}
T.~Theil, {\em {Physics Beyond the Standard Model at all Scales}}.
\newblock PhD thesis, Munich, Tech. U., 2023.

\bibitem{Misiak:1995}
M.~Misiak and M.~Münz, {\it Two-loop mixing of dimension-five flavor-changing
  operators},  {\em Physics Letters B} {\bf 344} (Jan., 1995) 308–318.

\bibitem{Chetyrkin:1997fm}
K.~G. Chetyrkin, M.~Misiak, and M.~Munz, {\it {Beta functions and anomalous
  dimensions up to three loops}},  {\em Nucl. Phys. B} {\bf 518} (1998)
  473--494, [\href{http://arxiv.org/abs/hep-ph/9711266}{{\tt hep-ph/9711266}}].

\bibitem{Zoller:2014}
M.~F. Zoller, {\it Three-loop beta function for the higgs self-coupling},
  2014.

\bibitem{Alloul:2013bka}
A.~Alloul, N.~D. Christensen, C.~Degrande, C.~Duhr, and B.~Fuks, {\it
  {FeynRules 2.0 - A complete toolbox for tree-level phenomenology}},  {\em
  Comput. Phys. Commun.} {\bf 185} (2014) 2250--2300,
  [\href{http://arxiv.org/abs/1310.1921}{{\tt arXiv:1310.1921}}].

\bibitem{Hahn:2000kx}
T.~Hahn, {\it {Generating Feynman diagrams and amplitudes with FeynArts 3}},
  {\em Comput. Phys. Commun.} {\bf 140} (2001) 418--431,
  [\href{http://arxiv.org/abs/hep-ph/0012260}{{\tt hep-ph/0012260}}].

\bibitem{Shtabovenko:2023idz}
V.~Shtabovenko, R.~Mertig, and F.~Orellana, {\it {FeynCalc 10: Do multiloop
  integrals dream of computer codes?}},
  \href{http://arxiv.org/abs/2312.14089}{{\tt arXiv:2312.14089}}.

\bibitem{Shtabovenko:2020gxv}
V.~Shtabovenko, R.~Mertig, and F.~Orellana, {\it {FeynCalc 9.3: New features
  and improvements}},  {\em Comput. Phys. Commun.} {\bf 256} (2020) 107478,
  [\href{http://arxiv.org/abs/2001.04407}{{\tt arXiv:2001.04407}}].

\bibitem{Shtabovenko:2016sxi}
V.~Shtabovenko, R.~Mertig, and F.~Orellana, {\it {New Developments in FeynCalc
  9.0}},  {\em Comput. Phys. Commun.} {\bf 207} (2016) 432--444,
  [\href{http://arxiv.org/abs/1601.01167}{{\tt arXiv:1601.01167}}].

\bibitem{Mertig:1990an}
R.~Mertig, M.~Bohm, and A.~Denner, {\it {FEYN CALC: Computer algebraic
  calculation of Feynman amplitudes}},  {\em Comput. Phys. Commun.} {\bf 64}
  (1991) 345--359.

\bibitem{Passarino:1978jh}
G.~Passarino and M.~J.~G. Veltman, {\it {One Loop Corrections for e+ e-
  Annihilation Into mu+ mu- in the Weinberg Model}},  {\em Nucl. Phys. B} {\bf
  160} (1979) 151--207.

\bibitem{Ellis:2007qk}
R.~K. Ellis and G.~Zanderighi, {\it {Scalar one-loop integrals for QCD}},  {\em
  JHEP} {\bf 02} (2008) 002, [\href{http://arxiv.org/abs/0712.1851}{{\tt
  arXiv:0712.1851}}].

\bibitem{Cyrol:2016zqb}
A.~K. Cyrol, M.~Mitter, and N.~Strodthoff, {\it {FormTracer - A Mathematica
  Tracing Package Using FORM}},  {\em Comput. Phys. Commun.} {\bf 219} (2017)
  346--352, [\href{http://arxiv.org/abs/1610.09331}{{\tt arXiv:1610.09331}}].

\bibitem{Laporta:2000dsw}
S.~Laporta, {\it {High-precision calculation of multiloop Feynman integrals by
  difference equations}},  {\em Int. J. Mod. Phys. A} {\bf 15} (2000)
  5087--5159, [\href{http://arxiv.org/abs/hep-ph/0102033}{{\tt
  hep-ph/0102033}}].

\bibitem{Chetyrkin:1981qh}
K.~G. Chetyrkin and F.~V. Tkachov, {\it {Integration by parts: The algorithm to
  calculate $\beta$-functions in 4 loops}},  {\em Nucl. Phys. B} {\bf 192}
  (1981) 159--204.

\bibitem{Maierhofer:2017gsa}
P.~Maierh\"ofer, J.~Usovitsch, and P.~Uwer, {\it {Kira\textemdash{}A Feynman
  integral reduction program}},  {\em Comput. Phys. Commun.} {\bf 230} (2018)
  99--112, [\href{http://arxiv.org/abs/1705.05610}{{\tt arXiv:1705.05610}}].

\bibitem{Moch:2005id}
S.~Moch, J.~A.~M. Vermaseren, and A.~Vogt, {\it {The Quark form-factor at
  higher orders}},  {\em JHEP} {\bf 08} (2005) 049,
  [\href{http://arxiv.org/abs/hep-ph/0507039}{{\tt hep-ph/0507039}}].

\bibitem{Becher:2006mr}
T.~Becher, M.~Neubert, and B.~D. Pecjak, {\it {Factorization and Momentum-Space
  Resummation in Deep-Inelastic Scattering}},  {\em JHEP} {\bf 01} (2007) 076,
  [\href{http://arxiv.org/abs/hep-ph/0607228}{{\tt hep-ph/0607228}}].

\bibitem{Beneke:2019vhz}
M.~Beneke, A.~Broggio, C.~Hasner, K.~Urban, and M.~Vollmann, {\it {Resummed
  photon spectrum from dark matter annihilation for intermediate and narrow
  energy resolution}},  {\em JHEP} {\bf 08} (2019) 103,
  [\href{http://arxiv.org/abs/1903.08702}{{\tt arXiv:1903.08702}}]. [Erratum:
  JHEP 07, 145 (2020)].

\bibitem{AH:2019pyp}
A.~A~H, P.~Mukherjee, and V.~Ravindran, {\it {Infrared structure of
  $SU(N)\times U(1)$ gauge theory to three loops}},  {\em JHEP} {\bf 08} (2020)
  156, [\href{http://arxiv.org/abs/1912.13386}{{\tt arXiv:1912.13386}}].

\bibitem{Luo:2002ey}
M.-x. Luo and Y.~Xiao, {\it {Two loop renormalization group equations in the
  standard model}},  {\em Phys. Rev. Lett.} {\bf 90} (2003) 011601,
  [\href{http://arxiv.org/abs/hep-ph/0207271}{{\tt hep-ph/0207271}}].

\bibitem{Bednyakov:2012rb}
A.~V. Bednyakov, A.~F. Pikelner, and V.~N. Velizhanin, {\it {Anomalous
  dimensions of gauge fields and gauge coupling beta-functions in the Standard
  Model at three loops}},  {\em JHEP} {\bf 01} (2013) 017,
  [\href{http://arxiv.org/abs/1210.6873}{{\tt arXiv:1210.6873}}].

\end{thebibliography}\endgroup

\end{document}